\pdfoutput=1
\documentclass[11pt,twoside,a4paper,cmspaper,final,collab]{cms-tdr}
\def\svnVersion{5fc2c1b}\def\svnDate{2022/10/22}\def\cmsCernNoTag{CERN-EP-2021-272}\def\cmsCernDate{\today}\def\cmsMessage{Published in Nature Physics as \href{http://dx.doi.org/10.1038/s41567-022-01682-0}{\doi{10.1038/s41567-022-01682-0}.}}
\begin{document}\cmsNoteHeader{HIG-21-013}

\newcommand{\fakesection}[1]{
  \par
  \ifthenelse{\boolean{cms@external}}{}{\renewcommand{\sectionmark}[1]{\markright{\textbf{#1}}}}
  \sectionmark{#1}
}
\newcommand{\beginsupplement}{
  \ifthenelse{\boolean{cms@external}}{
  \renewcommand{\figurename}{Extended Data Fig.}
  }
  {
  \renewcommand{\figurename}{Extended Data Figure}
  }
  \renewcommand{\tablename}{Extended Data Table}
  \numberwithin{table}{section}
  \numberwithin{figure}{section}
  \numberwithin{equation}{section}
  \setcounter{table}{0}
  \setcounter{figure}{0}
  \renewcommand{\thetable}{\arabic{table}}
  \renewcommand{\thefigure}{\arabic{figure}}
  \renewcommand{\theequation}{S\arabic{equation}}
}

\providecommand{\cmsTable}[1]{\ifthenelse{\boolean{cms@external}}{\resizebox{\columnwidth}{!}{#1}}{#1}}

\newcommand{\lumiC}{\ensuremath{2.3\fbinv{}}\xspace}
\newcommand{\lumiDE}{\ensuremath{78\fbinv{}}\xspace}
\newcommand{\lumiDEF}{\ensuremath{138\fbinv{}}\xspace}
\newcommand{\lumiCDEF}{\ensuremath{140\fbinv{}}\xspace}
\newcommand{\cPe}{\ensuremath{\Pe}\xspace}
\newcommand{\cPm}{\ensuremath{\PGm}\xspace}
\newcommand{\jhugen}{\textsc{jhugen}\xspace}
\newcommand{\mela}{\textsc{mela}\xspace}
\newcommand{\MelaAnalytics}{\textsc{MelaAnalytics}\xspace}
\newcommand{\pho}{\ensuremath{{\PGg}}\xspace}
\newcommand{\F}{\ensuremath{{\HepParticle{F}{}{}}}\xspace}
\newcommand{\X}{\ensuremath{{\HepParticle{X}{}{}}}\xspace}
\newcommand{\V}{\ensuremath{{\HepParticle{V}{}{}}}\xspace}
\newcommand{\W}{\ensuremath{{\PW}}\xspace}
\newcommand{\VV}{\ensuremath{\V{}\V{}}\xspace}
\newcommand{\WW}{\ensuremath{\W{}\W{}}\xspace}
\newcommand{\WZ}{\ensuremath{\W{}\PZ{}}\xspace}
\newcommand{\ZZ}{\ensuremath{\PZ{}\PZ{}}\xspace}
\newcommand{\ZGam}{\ensuremath{\PZ{}\pho{}}\xspace}
\newcommand{\ZllGam}{\ensuremath{\PZ(\to\ell\ell)\pho{}}\xspace}
\newcommand{\ZnunuGam}{\ensuremath{\PZ(\to\nu\nu)\pho{}}\xspace}
\newcommand{\WlnuGam}{\ensuremath{\PW(\to\ell\nu)\pho{}}\xspace}
\newcommand{\llGam}{\ensuremath{\ell\ell\pho{}}\xspace}
\newcommand{\HWW}{\ensuremath{\PH{}\PW{}\PW{}}\xspace}
\newcommand{\HZZ}{\ensuremath{\PH{}\PZ{}\PZ{}}\xspace}
\newcommand{\HVV}{\ensuremath{\PH{}\V{}\V{}}\xspace}
\newcommand{\VBF}{\ensuremath{\text{VBF}}\xspace}
\newcommand{\VH}{\ensuremath{\V{}\PH{}}\xspace}
\newcommand{\WH}{\ensuremath{\W{}\PH{}}\xspace}
\newcommand{\ZH}{\ensuremath{\PZ{}\PH{}}\xspace}
\newcommand{\qqZZ}{\ensuremath{\qqbar \to \ZZ}\xspace}
\newcommand{\PAQqpr}{{\HepAntiParticle{q}{}{\prime}}\xspace}
\newcommand{\qqbarprime}{\ensuremath{\PQq{}\PAQqpr}\xspace}
\newcommand{\qqWZ}{\ensuremath{\qqbarprime \to \WZ}\xspace}
\newcommand{\tZXtxt}{\ensuremath{\PQt{}\PZ{}}+\ensuremath{\X{}}\xspace}
\newcommand{\tVXtxt}{\ensuremath{\PQt{}\V{}}+\ensuremath{\X{}}\xspace}
\newcommand{\Zjetstxt}{\PZ{}+jets\xspace}
\newcommand{\Wjetstxt}{\PW{}+jets\xspace}
\newcommand{\Wlnujetstxt}{\ensuremath{\PW(\to\ell\nu)}+jets\xspace}
\newcommand{\Gamjetstxt}{\pho{}+jets\xspace}
\newcommand{\dphillmet}{\ensuremath{\Delta\phi^{\ell\ell}_{\text{miss}}}}
\newcommand{\dphilljetsmet}{\ensuremath{\Delta\phi^{\ell\ell\text{+jets}}_{\text{miss}}}}
\newcommand{\dphiWZjetsmet}{\ensuremath{\Delta\phi^{3\ell\text{+jets}}_{\text{miss}}}}
\newcommand{\mindphijetmet}{\ensuremath{\min{\Delta\phi^{\mathrm{j}}_{\text{miss}}}}}
\newcommand{\emu}{\ensuremath{\Pe\PGm}\xspace}
\newcommand{\Hboson}{\ensuremath{\PH} boson\xspace}
\newcommand{\mll}{\ensuremath{m_{\ell\ell}}\xspace}
\newcommand{\mTZZ}{\ensuremath{m_\mathrm{T}^{\ZZ{}}}\xspace}
\newcommand{\mTWZ}{\ensuremath{m_\mathrm{T}^{\WZ{}}}\xspace}
\newcommand{\mZZ}{\ensuremath{m_{\ZZ{}}}\xspace}
\newcommand{\mZ}{\ensuremath{m_{\PZ{}}}\xspace}
\newcommand{\mW}{\ensuremath{m_{\W{}}}\xspace}
\newcommand{\ptll}{\ensuremath{\pt^{\ell\ell}}\xspace}
\newcommand{\vecptll}{\ensuremath{\vec{p}_{\mathrm{T}}^{\,\ell\ell}}\xspace}
\newcommand{\vecptj}{\ensuremath{\vec{p}_{\mathrm{T}}^{\,\mathrm{j}}}\xspace}
\newcommand{\Nj}{\ensuremath{N_{\mathrm{j}}}\xspace}
\newcommand{\GH}{\ensuremath{\Gamma_{\PH{}}}\xspace}
\newcommand{\GHSM}{\ensuremath{\Gamma_{\PH{}}^{\text{SM}}}\xspace}
\newcommand{\mH}{\ensuremath{m_{\PH{}}}\xspace}
\newcommand{\mV}{\ensuremath{m_{\V{}}}\xspace}
\newcommand{\mVV}{\ensuremath{m_{\V{}\V{}}}\xspace}
\newcommand{\mell}{\ensuremath{m_{4\ell}}\xspace}
\newcommand{\glufu}{\ensuremath{\Pg{}\Pg{}}\xspace}
\newcommand{\muoffsh}{\ensuremath{\mu^{\text{off-shell}}}\xspace}
\newcommand{\mupoffsh}{\ensuremath{\mu^{\text{off-shell}}_{p}}\xspace}
\newcommand{\muFoffsh}{\ensuremath{\mu^{\text{off-shell}}_{\F{}}}\xspace}
\newcommand{\muVoffsh}{\ensuremath{\mu^{\text{off-shell}}_{\V{}}}\xspace}
\newcommand{\RVFoffsh}{\ensuremath{R^{\text{off-shell}}_{\V{},\F{}}}\xspace}
\newcommand{\muoffshshort}{
\ifthenelse{\boolean{cms@external}}{
\ensuremath{\mu^{\text{off.}}}\xspace
}{
\muoffsh
}
}
\newcommand{\muFoffshshort}{
\ifthenelse{\boolean{cms@external}}{
\ensuremath{\mu^{\text{off.}}_{\F{}}}\xspace
}{
\muFoffsh
}
}
\newcommand{\muVoffshshort}{
\ifthenelse{\boolean{cms@external}}{
\ensuremath{\mu^{\text{off.}}_{\V{}}}\xspace
}{
\muVoffsh
}
}
\newcommand{\RVFoffshshort}{
\ifthenelse{\boolean{cms@external}}{
\ensuremath{R^{\text{off.}}_{\V{},\F{}}}\xspace
}{
\RVFoffsh
}
}
\newcommand{\dNLL}{\ensuremath{-2\Delta\ln\mathcal{L}}\xspace}
\newcommand{\LC}[1]{\ensuremath{\Lambda_{#1}}\xspace}
\newcommand{\AC}[1]{\ensuremath{a_{#1}}\xspace}
\newcommand{\ai}{\ensuremath{\AC{i}}\xspace}
\newcommand{\fLC}[1]{\ensuremath{f_{\Lambda #1}}\xspace}
\newcommand{\fAC}[1]{\ensuremath{f_{a #1}}\xspace}
\newcommand{\fcospLC}[1]{\ensuremath{\fLC{#1}}\xspace}
\newcommand{\fcospAC}[1]{\ensuremath{\fAC{#1}}\xspace}
\newcommand{\fcospai}{\fcospAC{i}}
\newcommand{\offshell}{off-shell\xspace}
\newcommand{\onshell}{on-shell\xspace}
\newcommand{\pileup}{pileup\xspace}
\newcommand{\StandardModel}{standard model\xspace}
\newcommand{\BreitWigner}{Breit--Wigner\xspace}
\newcommand{\Dbkg}{\ensuremath{{\mathcal{D}}_{\text{bkg}}}\xspace}
\newcommand{\DjjVBF}{\ensuremath{{\mathcal{D}}^{{\VBF}}_{\text{2jet}}}\xspace}
\newcommand{\DjjVBFHS}{\ensuremath{{\mathcal{D}}^{{\VBF}, a2}_{\text{2jet}}}\xspace}

\cmsNoteHeader{HIG-21-013}
\title{Measurement of the Higgs boson width and evidence of its \offshell contributions to \ZZ production}
\titlerunning{Off-shell Higgs boson production}

\date{\today}

\abstract{
Since the discovery of the Higgs boson in 2012, detailed studies of its properties have been ongoing. Besides its mass, its width --- related to its lifetime --- is an important parameter. One way to determine this quantity is by measuring its \offshell production, where the Higgs boson mass is far away from its nominal value, and relating it to its \onshell production, where the mass is close to the nominal value. Here, we report evidence for such \offshell contributions to the production cross section of two $\PZ$ bosons with data from the CMS experiment at the CERN Large Hadron Collider. We constrain the total rate of the \offshell Higgs boson contribution beyond the $\PZ$ boson pair production threshold, relative to its \StandardModel expectation, to the interval $[0.0061, 2.0]$ at 95\% confidence level. The scenario with no \offshell contribution is excluded at a $p$-value of 0.0003 (3.6 standard deviations). We measure the width of the Higgs boson as $\Gamma_{\PH{}} = 3.2_{-1.7}^{+2.4}\MeV$, in agreement with the \StandardModel expectation of $4.1\MeV$. In addition, we set constraints on anomalous Higgs boson couplings to $\PW$ and $\PZ$ boson pairs.
}

\hypersetup{
pdfauthor={CMS Collaboration},
pdftitle={Measurement of the Higgs boson width and evidence of its off-shell contributions to ZZ production},
pdfsubject={CMS},
pdfkeywords={Higgs, off-shell, width, anomalous coupling, HVV, EFT}
}

\maketitle

The \StandardModel (SM) of particle physics provides an elegant description for the masses and interactions of fundamental particles.
These are fermions, which are the building blocks of ordinary matter, and gauge bosons, which are the carriers of the electroweak (EW) and strong forces.
In addition, the SM postulates the existence of a quantum field responsible for the
generation of the masses of fundamental particles through a phenomenon known as the Brout--Englert--Higgs mechanism. This field, known as the \emph{Higgs field}~\cite{Englert:1964et,Higgs:1964pj,Guralnik:1964eu}, interacts with SM particles, giving them mass, as well as with itself.
The field carrier is a massive, scalar (spin-0) particle known as the Higgs ($\PH$) boson.
Nearly half a century after its postulation, it was finally observed in 2012 with a mass \mH of around 125\GeV by the ATLAS and CMS
Collaborations~\cite{Aad:2012tfa,Chatrchyan:2012xdj,Chatrchyan:2013lba} at the CERN Large Hadron Collider (LHC). Given the unique role
the \Hboson plays in the SM, studies of its properties are a major goal of particle physics.

Apart from mass, another important property of a particle is its lifetime $\tau$.
Only a few fundamental particles are stable; others---including
the {\Hboson{}}---exist only for a fleeting moment before disintegrating into other, lighter, species.
The Heisenberg uncertainty principle~\cite{Heisenberg:1927} provides a direct connection between the lifetime of a particle and the uncertainty in its mass, a property known as the particle's width, $\Gamma$.
Any unstable particle (often referred to as a \emph{resonance}) has a finite lifetime,
with shorter $\tau$ corresponding to broader $\Gamma$.
The two quantities are related through the Planck constant, $h$, as $\Gamma=h / (2 \pi \tau)$.
Even with perfect experimental resolution, the observed mass of an unstable particle will not be constant across a series of measurements (\eg, of the invariant mass of its decay products $i$, which is calculated from the sums of their energies, $E_i$, and momenta, $\vec{p}_i$, as $\sqrt{(\sum_{i}{E_i})^2 - \abs{\sum_{i}{\vec{p}_i}}^2}$).
The possible mass values are distributed according to a characteristic relativistic \BreitWigner distribution~\cite{RelBreitWigner}
with a nominal mass value corresponding to the maximum of the \BreitWigner{}, and with width parameter $\Gamma$.

Particles are understood to be \emph{on the mass shell} ({\onshell{}}) if their mass is close to the nominal mass value, and \emph{\offshell{}} if their mass takes a value far away from it.
By the aforementioned property of the \BreitWigner line shape, particles are generally more 
likely to be produced \onshell than \offshell when energy and momentum conservation allows it.
Scattering amplitudes ($\mathrm{A}$) for \offshell particle production, followed by a specific decay final state, may be modified further by interference with other processes, which is large and destructive in the case of the \Hboson{}. In this specific case, writing $\mathrm{A}=\PH+\mathrm{C}$, with $\PH$ standing for the \Hboson contribution and $\mathrm{C}$ for other interfering contributions, we will use the term ``\offshell production''
as a shorthand for the $\abs{\PH{}}^2$ term in $\abs{\mathrm{A}}^2$.

For broad resonances, the width can be obtained by directly measuring the \BreitWigner line shape, \eg, as was done in the case of 
the \PZ boson, measured to have a mass of $\mZ=91.188 \pm 0.002\GeV$ and a width of $\Gamma_\PZ = 2.495 \pm 0.002\GeV$ at the CERN Large Electron Positron collider~\cite{ALEPH:2005ab}.
The \Hboson is expected to live three orders of magnitude longer, with a theoretically predicted width of $\GH = 4.1\MeV$ ($0.0041\GeV$)~\cite{deFlorian:2016spz}, and a deviation from the SM prediction would indicate the existence of new physics. This width is too small to be measured directly from the line shape
because of the limited mass resolution of order $1\GeV$ achievable with the present LHC detectors.
Another direct way of measuring the \Hboson width would be 
to measure its lifetime by means of its decay length and use the relationship $\GH=h / (2 \pi \tau_\PH)$,
but its lifetime is still too short ($\tau_\PH = 1.6 \times 10^{-22}\unit{s}$) to be detectable directly. The present experimental limit on this quantity is $\tau_\PH < 1.9 \times 10^{-13}\unit{s}$ at $95\%$ confidence level ({\CL})~\cite{Khachatryan:2015mma}, nine orders of magnitude above the SM lifetime.

The value of $\GH$ can be extracted with much better precision through a combined measurement of \onshell and \offshell \Hboson production.
In the decay of an
\Hboson with $\mH \approx 125\GeV$ to a pair of massive gauge bosons \V ($\V=\PW$ or $\PZ$, with masses around $80.4$ or $91.2\GeV$, respectively), we have $\mV<\mH<2\mV$.
Therefore, when the \Hboson is produced \onshell (with the \VV invariant mass $\mVV \sim \mH$), one of the \V bosons must be \offshell{} to
satisfy four-momentum conservation.
Once the \Hboson is produced \offshell with large enough invariant mass $\mVV>2\mV$ (\offshell \Hboson production region), the \V bosons themselves are produced {\onshell}.
Since the \BreitWigner mass distribution of either the $\PH$ or $\V$ boson maximizes at their respective nominal masses, the rate of \offshell \Hboson production above the
\V boson pair production threshold is enhanced with respect to what one would expect from the \BreitWigner line shape of the \Hboson alone.

The measurement of the higher part of the $\mVV$ spectrum can then be used to establish \offshell \Hboson production.
The ratio of \offshell to \onshell production rates allows for a measurement of $\GH$~\cite{Caola:2013yja,Campbell:2013una}
via the cross section proportionality relations
\begin{equation*}
\sigma^{\text{\onshell{}}} \propto \frac{g_p^2 g_d^2}{\GH} \propto \mu_p
\Rightarrow
\sigma^{\text{\offshell{}}} \propto g_p^2 g_d^2 \propto \mu_p\, \GH{},
\end{equation*}
where $g_p$ and $g_d$ are the couplings associated with the \Hboson production and decay modes, respectively, and $\mu_p$ is the \onshell \Hboson signal strength in the production mode being considered.
Each signal strength is defined as the ratio of the \Hboson squared amplitude in the measured cross section to that predicted in the SM.
The \offshell \Hboson signal strength, $\mupoffsh$, can be expressed as $\mu_p\, \GH$ in each production mode,
and the scenario with no \offshell production becomes equivalent to the limiting case $\GH=0$.
For the rest of this article, we concentrate on the \ZZ decay channel, \ie, $g_d$ corresponding to the $\PH \to \ZZ$ decay.
The CMS and ATLAS Collaborations have previously used this method to set upper limits on $\GH$ as low as $9.2\MeV$ at $95\%$ \CL~\cite{Aaboud:2018puo,Sirunyan:2019twz}.

\begin{figure*}[htbp]
\centering
\includegraphics[width=\textwidth]{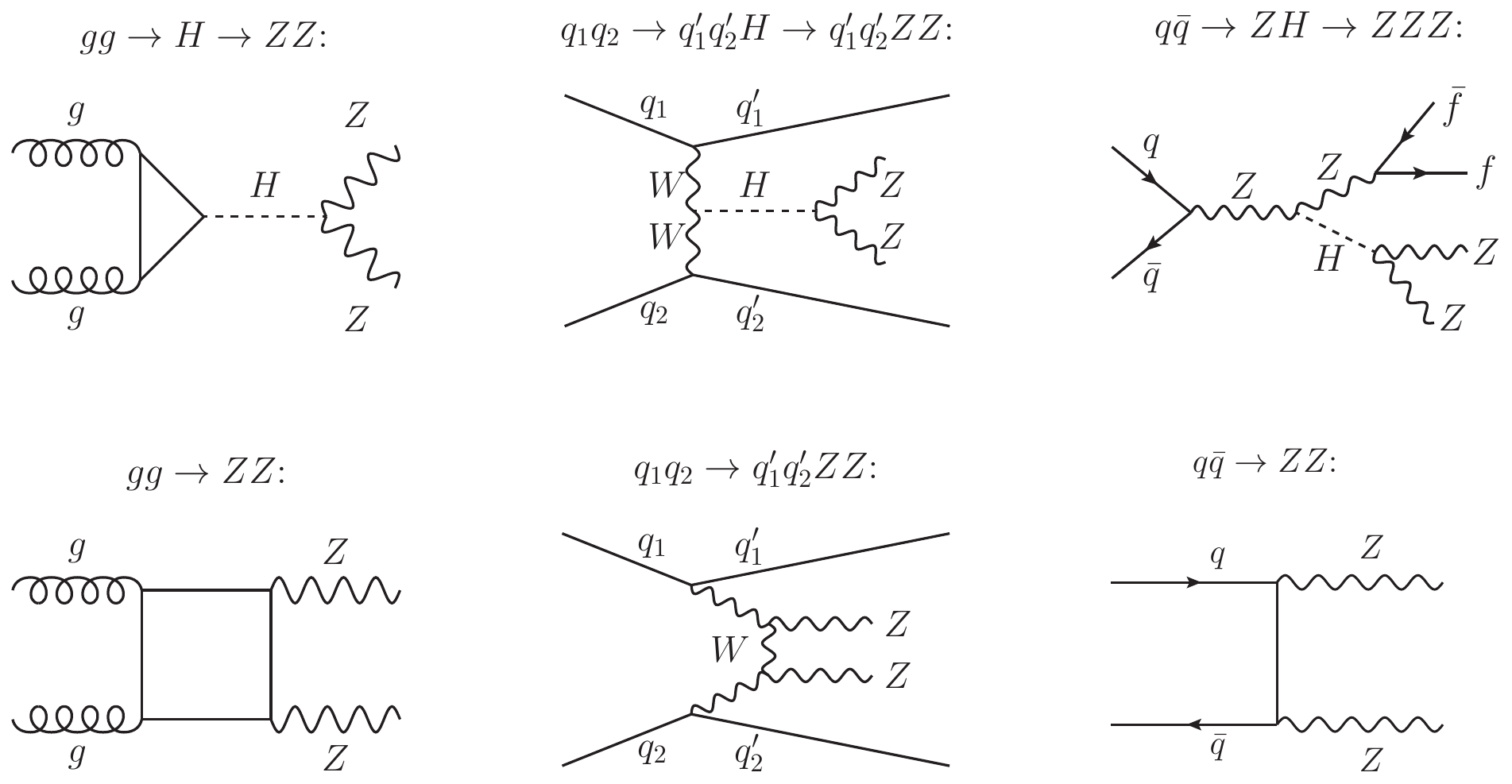}
\caption{
\textbf{Feynman diagrams for important contributions to \ZZ production.}
Diagrams can be distinguished as those involving the \Hboson (top), and those that give rise to continuum \ZZ production (bottom).
The interaction displayed at tree level in each diagram is meant to progress from left to right. Each straight, curvy, or curly line refers to the different set of particles denoted. Straight, solid lines with no arrows indicate the line could refer to either a particle or an antiparticle, whereas those with forward (backward) arrows refer to a particle (an antiparticle).
}
\label{fig:feyngg}
\end{figure*}

It is important to distinguish between two types of \Hboson production modes:
the gluon fusion $\glufu\to\PH\to\ZZ$ process, where the \Hboson is produced via its couplings to fermions, and the
EW processes, which involve \HVV (\ie, $\HWW$ or $\HZZ$) couplings. The top row of Fig.~\ref{fig:feyngg} shows
the Feynman diagrams for the most dominant
contributions to the $\glufu$ (top left) process, and the EW processes of
vector boson fusion ($\VBF$, top center) and $\VH$ (top right).  
A more complete set of diagrams for the EW process
are shown in Extended Data Figs.~\ref{fig:supp:feyn-EWsig} and \ref{fig:supp:feyn-EWcontin}.
Because different \Hboson couplings are involved in the $\glufu$ and EW processes, we extract two
\offshell signal strength parameters $\muFoffsh$ for the $\glufu$ mode
and $\muVoffsh$ for the EW mode.
We also consider an overall \offshell signal strength parameter $\muoffsh$ with different assumptions on the ratio
$\RVFoffsh=\muVoffsh/\muFoffsh$.

A major challenge
arises from the fact that there are other sources of \ZZ pairs in the SM (continuum \ZZ production), see for example
the bottom row of Fig.~\ref{fig:feyngg}.
These contributions, particularly those from
$\qqZZ$, are typically much larger than the contribution from off-shell $\PH \to \ZZ$.
In addition, some of the amplitudes from continuum \ZZ processes interfere with the
\Hboson amplitudes because they share the same initial and final states.
For example, the amplitudes in the first column of Fig.~\ref{fig:feyngg}, or those in the second column,
interfere with each other; the amplitude shown in the lower right panel (shown more generically in Extended Data Fig.~\ref{fig:supp:feyn-qqVZ}) does not interfere with any of the other diagrams as we omit the negligible contribution of $\qqbar \to \PH \to \ZZ$ that would interfere with it.

\begin{figure*}[htbp]
\centering
\includegraphics[width=.45\textwidth]{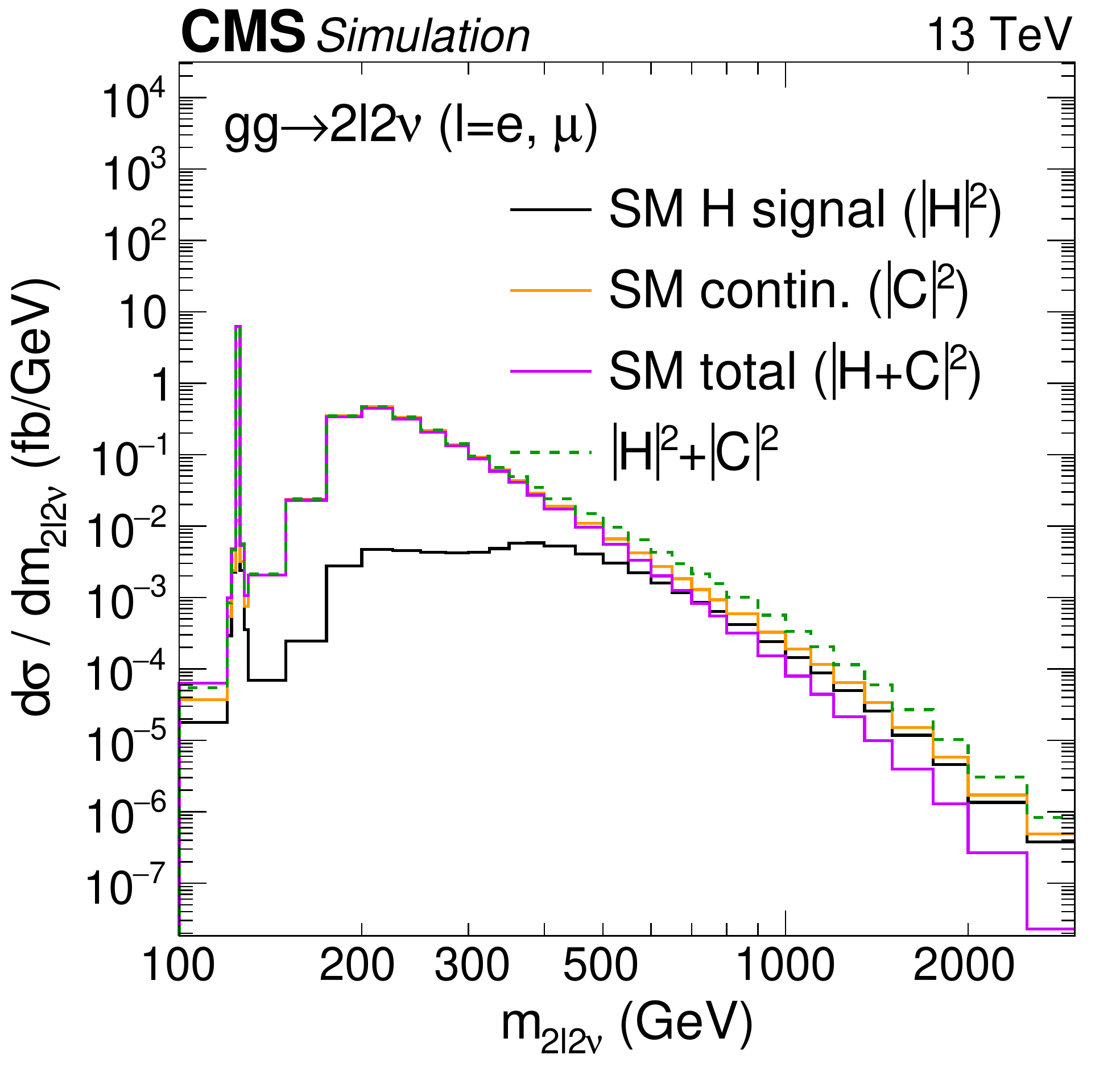}
\includegraphics[width=.45\textwidth]{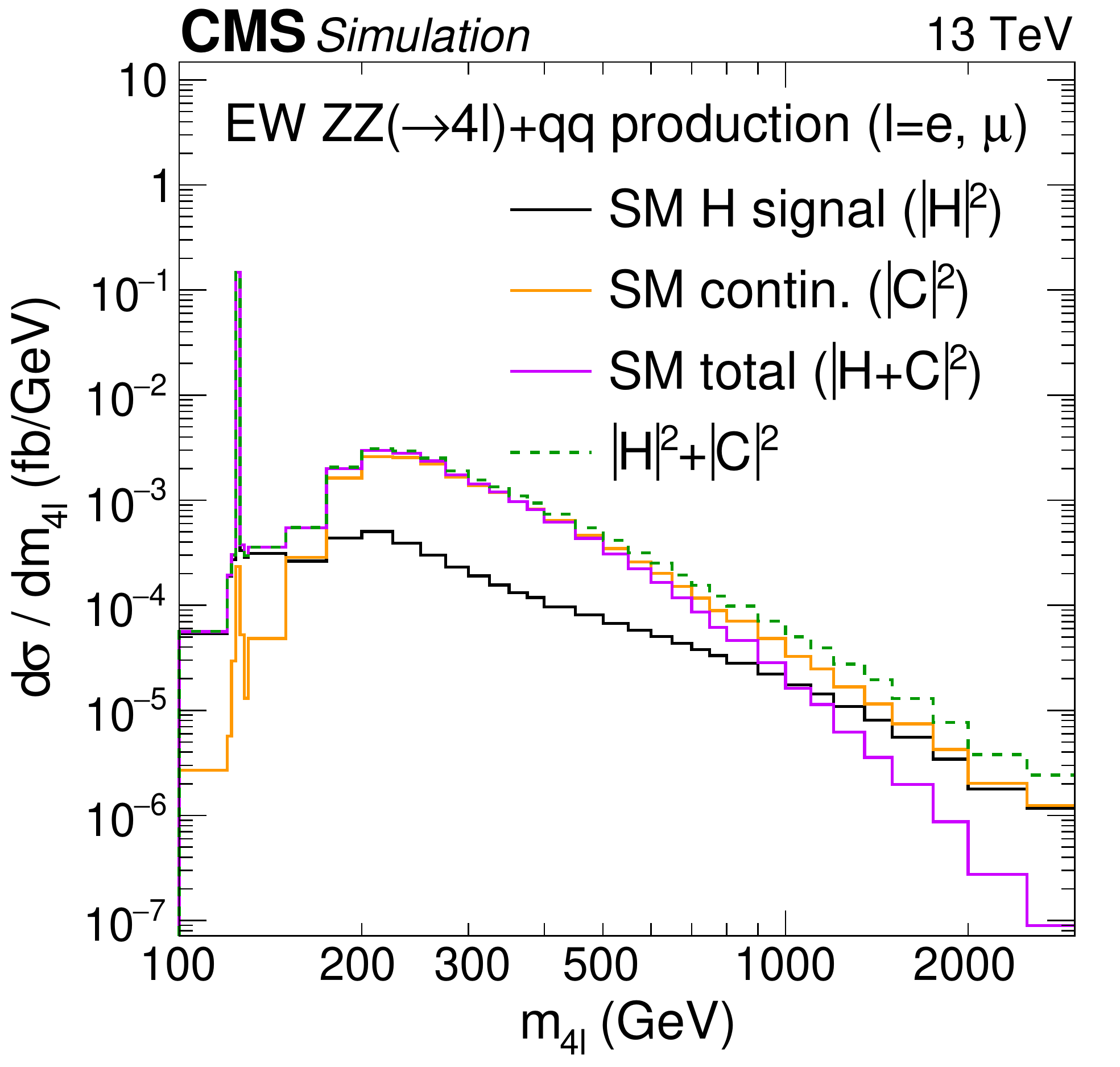}
\caption{
\textbf{Standard model calculations of $\ZZ$ invariant mass in the \glufu and EW processes.}
  Shown are the distributions for the $2\ell2\nu$ invariant mass, $m_{2\ell2\nu}$, from the $\glufu \to 2\ell2\nu$ process on the left panel, and the $4\ell$ invariant mass, $\mell$, from the EW $\ZZ (\to 4\ell)+\PQq{}\PQq{}$ processes on the right.
  These processes involve the \Hboson ($\abs{\PH{}}^2$) and interfering continuum ($\abs{\mathrm{C}}^2$) contributions to the scattering amplitude, shown in black and gold, respectively.
  The dashed green curve represents their direct sum without the interference ($\abs{\PH{}}^2+\abs{\mathrm{C}}^2$), and the solid magenta curve represents the sum with interference included ($\abs{\PH{}+\mathrm{C}}^2$).
  Note that the interference is destructive, and its importance grows as the mass increases.
  The integrated luminosity is taken to be $1\fbinv$, so these distributions are equivalent to the differential cross section spectra $d\sigma/dm_{2\ell2\nu}$ (left) and $d\sigma/d\mell$ (right).
  The distributions are shown after requiring that all charged leptons satisfy $\pt>7\GeV$ and $\abs{\eta}<2.4$, and that the invariant mass of any charged lepton pair with same flavor and opposite charge is greater than $4\GeV$.
  Here, $\pt$ denotes the magnitude of the momentum of these leptons transverse to the $\Pp{}\Pp{}$ collision axis, and $\eta$ denotes their pseudorapidity, defined as $-\ln\left[\tan\left(\theta/2\right)\right]$ using the angle $\theta$ between their momentum vector and the collision axis.
  Calculations for the $\glufu \to 4\ell$ and EW $\ZZ (\to 2\ell2\nu)+\PQq{}\PQq{}$ processes exhibit similar
  qualitative properties.
  The details of the Monte Carlo programs used for these calculations are given in the Methods section.
}
\label{fig:mZZ-ggEW}
\end{figure*}

The interference between the \Hboson and continuum \ZZ amplitudes is destructive~\cite{LlewellynSmith:1973ey,Cornwall:1974km,Lee:1977eg,Glover:1988rg,Campanario:2012bh,Kauer:2012hd}.
This destructive interference plays a key role in the SM as it is one of the contributions that unitarizes the scattering of massive gauge bosons, keeping the computation of the cross section for \ZZ production in proton-proton ($\Pp\Pp$) collisions finite~\cite{LlewellynSmith:1973ey,Cornwall:1974km,Lee:1977eg,Glover:1988rg}.
Figure~\ref{fig:mZZ-ggEW} displays the interplay between the \Hboson production modes and the interfering continuum amplitudes, illustrating the growing importance of their destructive interference as \mZZ grows in the two final states included in the analysis, $\ZZ \to 2\ell2\nu$ and $\ZZ \to 4\ell$.
In the parametrization of the total cross section, contributions from this type of interference between the \Hboson and continuum \ZZ amplitudes scale with $\sqrt{\smash[b]{\muFoffsh}}$ and $\sqrt{\smash[b]{\muVoffsh}}$ for the $\glufu$ and EW modes, respectively.

In this article, we study  
\offshell \Hboson decays to $\ZZ \to 2\ell2\nu$, and
\onshell as well as \offshell \Hboson decays to $\ZZ \to 4\ell$
($\ell=\cPm$ or $\cPe$),
using a sample of $\Pp{}\Pp{}$
collisions at 13\TeV
collected by the CMS experiment at the LHC.
The selection and analysis of the \offshell
$\PZ\PZ \to 2\ell2\nu$ data sample is described in detail in this
article, and it is based on data collected between 2016 and 2018,
corresponding to an integrated luminosity of $\lumiDEF$.
For the $\PZ\PZ \to 4\ell$ mode, we use
previously published CMS
\offshell
(2016 and 2017 data sets, $\lumiDE$~\cite{Sirunyan:2019twz}) and
\onshell 
(2015~\cite{Sirunyan:2017tqd,Sirunyan:2019twz} and 2016--2018~\cite{Sirunyan:2021fpv} data sets, $\lumiC$ and $\lumiDEF$, respectively)
results.

Information on the \offshell signal strengths, $\GH$, and constraints on
possible beyond-the-SM (BSM)
anomalous couplings
are extracted from combined fits over several
kinematic distributions of the selected $2\ell2\nu$ and $4\ell$ events.
While \offshell events are the ones solely used to establish the presence of
\offshell \Hboson production,
the measurement of $\GH$ relies on the combination
of \onshell and \offshell data.

Because of the
presence of neutrinos, the \Hboson mass cannot be precisely
reconstructed in the $\PH \to 2\ell2\nu$ final state
as the longitudinal component of the total momentum carrried by the neutrinos cannot be measured.
Thus, \onshell
information can only be extracted from the $4\ell$ mode.
This combination of $4\ell$ and $2\ell2\nu$ data enables
the measurement of $\GH$ with a precision
of ${\sim}50\%$.
The measurement improves the upper limit on
$\tau_\PH$ by eight orders of magnitude compared to the direct
constraint from Ref.~\cite{Khachatryan:2015mma}.
The inclusion of the $2\ell2\nu$ data also allows the lower limits on $\muVoffsh$ to reach within ${\sim}65\%$ of its best fit value,
compared to the weaker constraints from $4\ell$ data alone, which reach within ${\sim}90\%$ of the $4\ell$-only best fit value~\cite{Sirunyan:2019twz}.

The \mZZ line shape is sensitive to the potential presence of anomalous \HVV couplings~\cite{Gainer:2014hha,Englert:2014aca,Khachatryan:2015mma,deFlorian:2016spz,Sirunyan:2019twz,Gritsan:2020pib}.
Thus, BSM physics could affect the ratio of \offshell to \onshell \Hboson production rates, and therefore
the measurement of $\GH$.
We test the effect of these couplings on the $\GH$ measurement and constrain the contribution from these couplings themselves.
In parametrizing anomalous \HVV contributions,
we adopt the formalism of Ref.~\cite{Sirunyan:2019twz}
with the scattering amplitude
\ifthenelse{\boolean{cms@external}}{
\begin{linenomath}
\begin{equation*}
\begin{aligned}
\mathrm{A} &\propto \left[ \AC{1} - \frac{q_{1}^2 + q_{2}^{2}}{\LC{1}^{2}} \right]
\times \mV^2 \epsilon_{1}^* \epsilon_{2}^* \\
&+ \AC{2} f_{\mu \nu}^{*(1)}f^{*(2)\,\mu\nu}
+ \AC{3} f^{*(1)}_{\mu \nu} {\tilde f}^{*(2)\,\mu\nu}.
\end{aligned}
\end{equation*}
\end{linenomath}
}{
\begin{equation*}
\mathrm{A} \propto\left[ \AC{1} - \frac{q_{1}^2 + q_{2}^{2}}{\LC{1}^{2}} \right]
\mV^2 \epsilon_{1}^* \epsilon_{2}^*
+ \AC{2} f_{\mu \nu}^{*(1)}f^{*(2)\,\mu\nu}
+ \AC{3} f^{*(1)}_{\mu \nu} {\tilde f}^{*(2)\,\mu\nu}.
\end{equation*}
}
Here, the polarization vector (four-momentum) of the vector boson $\V_i$ is denoted by $\epsilon_{i}$ ($q_i$) while $f^{(i){\mu \nu}} = \epsilon_{i}^{\mu}q_{i}^{\nu} - \epsilon_{i}^\nu q_{i}^{\mu}$ and ${\tilde f}^{(i)}_{\mu \nu} = \frac{1}{2} \epsilon_{\mu\nu\rho\alpha} f^{(i)\,\rho\alpha}$ are tensor expressions for each $\V_i$.
The BSM couplings
$\AC{2}$, $\AC{3}$, and $1/\LC{1}^{2}$ (denoted generically as $\ai$) are assumed to be real and can take negative values, with
the $\kappa$ factors in Ref.~\cite{Sirunyan:2019twz} absorbed into the definition of $1/\LC{1}^{2}$.
The first two are coefficients for generic CP-conserving and
CP-violating higher dimensional operators, respectively, while
$1/\LC{1}^{2}$ is the coefficient for the first-order term in the expansion of a SM-like tensor structure
with an anomalous dipole form factor in the invariant masses of the two \V bosons.
In what follows, we will use the shorthand ``$\ai$ hypothesis'' to refer
to the scenario where all BSM \HVV couplings other than $\ai$ itself are zero.

Throughout this work, we assume that
the gluon fusion loop amplitudes do not receive new physics
contributions apart from a rescaling of the SM amplitude.
Possible modifications of the \mZZ line shape~\cite{Sarica:2019dsz,Gritsan:2020pib} are neglected based on
existing LHC constraints~\cite{CMS:2020cga,ATLAS:2020ior,Ethier:2021bye}.

\subsection*{\texorpdfstring{$2\ell2\nu$}{2l2nu} analysis considerations}

The $2\ell2\nu$ analysis is based on the reconstruction of
$\PZ \to \ell\ell$ decays with a second
$\PZ$ boson decaying to neutrinos that escape detection.
The momentum of the undetected $\PZ$ boson transverse to the $\Pp{}\Pp{}$ collision axis can be measured through an imbalance across all remaining particles, \ie, missing transverse momentum ($\ptmiss$ or
\ptvecmiss in vector form). Thus, the analysis requires
large $\ptmiss$ as the $\PZ \to \nu \nu$ signature.

The event selection is sensitive to the tail of the instrumental \ptmiss resolution
in $\Pp{}\Pp{} \to$~\Zjetstxt events that constitute an important reducible background.
This contribution is estimated through a study of a data control
region (CR) of \Gamjetstxt events, where \ptmiss is purely instrumental as it is in \Zjetstxt events.

Processes such as $\Pp\Pp \to \ttbar$ or $\WW$ result
in nonresonant dilepton final states of same ($\Pep\Pem$ and $\PGmp\PGmm$) and opposite
flavor ($\Pepm\PGmmp$) with the same
probability and the same kinematic properties.
Thus, their background contribution to the $2\ell2\nu$ signal,
which includes two leptons of the same flavor, is estimated from an
opposite-flavor $\Pe\PGm$ CR.

Other backgrounds from
$\qqZZ$, $\qqWZ$ with $\PW \to \ell \nu$ and an undetected lepton, and
the small contribution from $\PQt\PZ$ production
are estimated from simulation.
A third CR of trilepton events, consisting mostly of
$\qqWZ$ events, is used to constrain the $\qqWZ$ background 
and, most importantly, the large
$\qqZZ$ background.
The ability to constrain $\qqZZ$ from
$\qqWZ$ is based on the similarity in the physics of these
processes.

Further details on event selection, kinematic observables,
and the methods to estimate the different contributions 
are discussed in the Methods section.

\subsection*{\texorpdfstring{$2\ell2\nu$}{2l2nu} kinematic observables}

The analysis of \offshell \Hboson events is based on $\mZZ$.
This quantity is computed from the reconstructed momenta
in the $4\ell$ final state as the invariant mass of the $4\ell$ system, $\mell$.
However, because of the undetected neutrinos,
we can only use the transverse mass \mTZZ{}, defined below,
as a proxy for $\mZZ$ in the $2\ell2\nu$ final state.
First, we identify \ptvecmiss as the transverse momentum vector
of the \PZ boson decaying into neutrinos.
Since there is no information on the longitudinal momenta
of the neutrinos, \mTZZ
is then computed as the invariant mass of the \ZZ pair with all longitudinal momenta set to zero.
This results in a variable with a distribution
that peaks at $\mZZ$, with a long tail towards lower values.
The definition of \mTZZ is
\ifthenelse{\boolean{cms@external}}{
\begin{linenomath}
\begin{equation*}
\begin{aligned}
{( \mTZZ )}^2 &=
\Biggl[ \sqrt{{\ptll}^2+{\mll}^2} + \sqrt{{\ptmiss}^2+{\mZ}^2} \Biggr]^2 \\
 &- {\bigg\lvert \vecptll + \ptvecmiss \bigg\rvert}^2,
\end{aligned}
\label{eq:mTZZ}
\end{equation*}
\end{linenomath}
}{
\begin{equation*}
{\left( \mTZZ \right)}^2 =
\left[ \sqrt{{\ptll}^2+{\mll}^2} + \sqrt{{\ptmiss}^2+{\mZ}^2} \right]^2
 - {\left| \vecptll + \ptvecmiss \right|}^2,
\label{eq:mTZZ}
\end{equation*}
}
where $\vecptll$ and \mll are the dilepton transverse momentum and invariant mass,
respectively, and \mZ, the \PZ boson pole mass, is taken to be $91.2\GeV$.

The kinematic quantity \ptmiss itself is used as another observable
to discriminate processes with genuine, large \ptmiss against the \Zjetstxt background.
Finally, in events with at least two jets,
we use matrix element (\mela~\cite{Gritsan:2020pib}) kinematic discriminants
that distinguish the \VBF process from the \glufu process or SM backgrounds.
These discriminants are the $\DjjVBF$-type kinematic discriminants used in Refs.~\cite{Sirunyan:2019twz,Sirunyan:2021fpv}, and 
are based on the
four-momenta of the \Hboson and the two jets leading in $\pt$.

\subsection*{Data interpretation}

The results for the \offshell signal strength parameters \muFoffsh{}, \muVoffsh{}, and \muoffsh, and the \Hboson width $\GH$
are extracted from binned extended maximum likelihood 
fits over several kinematic distributions following the parametrization in Ref.~\cite{Sirunyan:2019twz}.
In this parametrization, all mass dependencies are absorbed into the distributions for the various terms contributing to the likelihood, and the \offshell signal strength parameters, or $\GH$, are kept mass-independent.
Over different data periods and event categories, 117 multidimensional distributions are used in the fit:
42 for \offshell $2\ell2\nu$ data (10\,867 events), including 18 distributions from the trilepton \WZ CR (8541 events), and
18 and 57 for \offshell and \onshell $4\ell$ data (1407 \offshell and 621 \onshell events), respectively.

In the $2\ell2\nu$ data sample,
the value of \mTZZ is required to be greater than $300\GeV$.
Depending on the number of jets ($\Nj$),
this sample
is binned in
\mTZZ{} and \ptmiss{} ($\Nj<2$),
or \mTZZ{}, \ptmiss{},
and the $\DjjVBF$-type kinematic discriminants ($\Nj\geq2$).
For the $4\ell$ samples, the binning is in $\mell$ and \mela
discriminants, which are sensitive 
to differences between the \Hboson signal and continuum \ZZ production, or
the interfering amplitudes, or anomalous \HVV couplings.
These variables are listed in 
Table II of Ref.~\cite{Sirunyan:2019twz} for $4\ell$ \offshell data,
under `Scheme 2' in Table IV of Ref.~\cite{Sirunyan:2021fpv} for \onshell 2016-2018 data, and
in Table 1 of Ref.~\cite{Sirunyan:2019twz} for \onshell 2015 data.
The $\mell$ range is required to be within $105$--$140\GeV$ for $4\ell$ \onshell data,
or above $220\GeV$ for $4\ell$ \offshell data.

Theoretical uncertainties in the kinematic
distributions include the simulation of extra jets
(up to 20\% depending on $\Nj$),
and the quantum chromodynamic (QCD) running scale
and parton distribution function (PDF) uncertainties
in the cross section calculation (up to
30\% and 20\%, respectively, depending on the process, and
\mTZZ or \mell{}).
These are particularly important in the \glufu process since it
cannot be constrained by the trilepton \WZ CR.
Theory uncertainties also include those associated with the EW
corrections to the $\qqbar \to \ZZ$ and $\WZ$
processes, which reach 20\% at masses around $1\TeV$~\cite{Bierweiler:2013dja,Manohar:2016nzj}.

Experimental uncertainties include uncertainties in the
lepton reconstruction and trigger efficiency (typically 1\% per lepton),
the integrated luminosity (between 1.2\% and 2.5\%,
depending on the data-taking period~\cite{CMS-LUM-17-003,CMS-PAS-LUM-17-004,CMS-PAS-LUM-18-002}), and
the jet energy scale and resolution~\cite{Khachatryan:2016kdb},
which affect the
counting of jets, as well as the reconstruction of
the \VBF discriminants.

\subsection*{Evidence for \offshell contributions, and width measurement}

\begin{figure*}[htb!p]
\centering
\includegraphics[width=0.45\textwidth]{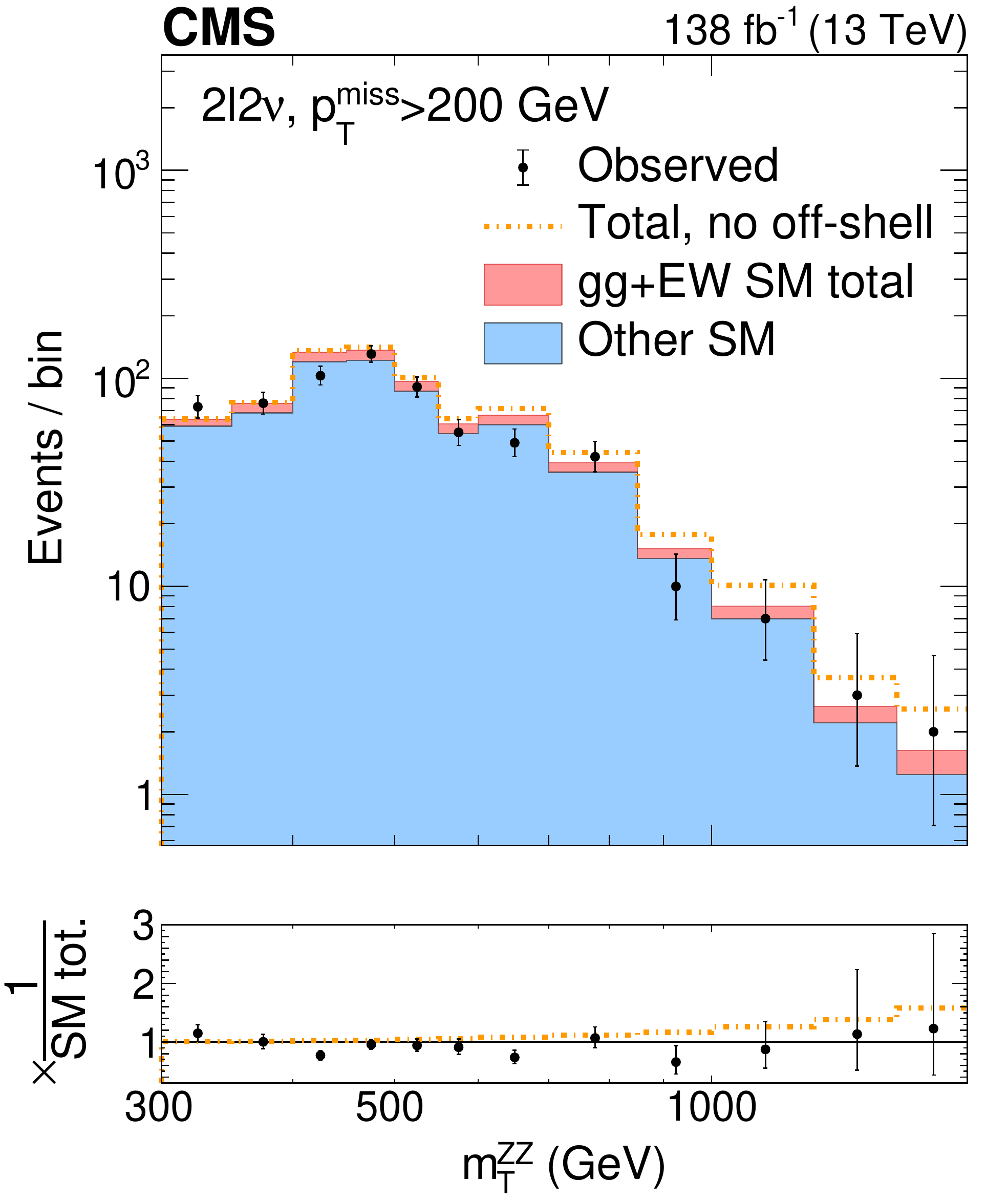}
\includegraphics[width=0.45\textwidth]{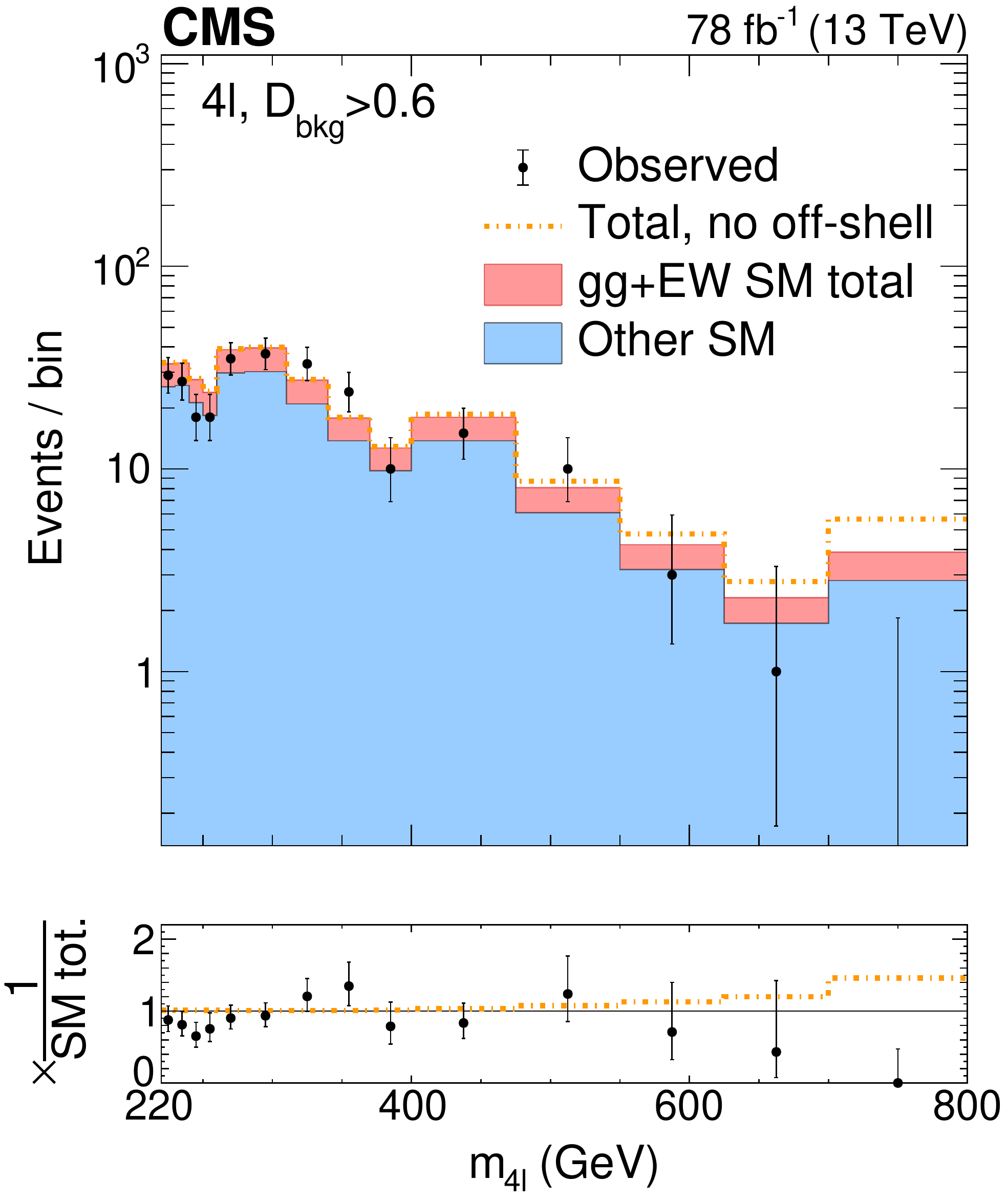}
\caption{
\textbf{Distributions of \ZZ invariant mass observables in the \offshell signal regions.}
The distributions of transverse \ZZ invariant mass, $\mTZZ$ from the $2\ell2\nu$ \offshell signal region are displayed on the left panel, and those of the $4\ell$ invariant mass, $\mell$, from the $4\ell$ \offshell signal region are displayed on the right.
The stacked histogram displays the distribution after a fit to the data with SM couplings,
with the blue filled area corresponding to the SM processes that do not include \Hboson interactions,
and the pink filled area adding processes that include \Hboson and interference contributions.
The gold dot-dashed line shows the fit to the no \offshell hypothesis.
The black points with error bars as uncertainties at 68\%~\CL show the observed data, which is consistent with the prediction with SM couplings within one standard deviation.
The last bins contain the overflow.
The requirements on the missing transverse momentum $\ptmiss$ in $2\ell2\nu$ events, and the $\Dbkg$-type kinematic background discriminants (see Table II of Ref.~\cite{Sirunyan:2019twz}) in $4\ell$ events are applied in order to enhance the \Hboson signal contribution.
The values of integrated luminosity displayed correspond to those included in the \offshell analyses of each final state.
The bottom panels show the ratio of the data or dashed histograms to the SM prediction (stacked histogram). The black horizontal line in these panels marks unit ratio.
}
\label{fig:mZZ}
\end{figure*}

A representative distribution of \mTZZ{}, integrated over all $\Nj$, is shown for $2\ell2\nu$ events on the left panel of Fig.~\ref{fig:mZZ}.
Finer details in terms of $\Nj$ and the various contributions to the event sample are
displayed in Extended Data Fig.~\ref{fig:supp:mTZZ-SR}.
The CRs for instrumental \ptmiss and nonresonant dilepton production backgrounds are illustrated in Extended Data Figs.~\ref{fig:supp:InstrMET} and \ref{fig:supp:NRB}, respectively, and the CR with trilepton \WZ events is illustrated in Extended Data Fig.~\ref{fig:supp:WZCR}. 
Also shown on the right panel of Fig.~\ref{fig:mZZ} is a representative distribution of $\mell$ from the combined \offshell $4\ell$ events.

The constraints on \muFoffsh{}, \muVoffsh{}, \muoffsh{}, and \GH
are summarized in Table~\ref{table:interp}, where we show the ``observed'' results, \ie, those extracted from data, as well as the ``expected''
ones, \ie, those based on the SM and our understanding of selection efficiencies, backgrounds, and 
systematic uncertainties.
The two set of results are consistent with statistical fluctuations in the data.
The constraint on $\GH$ at $95\%$ confidence level corresponds to
$7.7 \times 10^{-23}<\tau_\PH<1.3 \times 10^{-21}\unit{s}$ in \Hboson lifetime.

\begin{figure*}[htb!]
\centering
\includegraphics[width=0.45\textwidth]{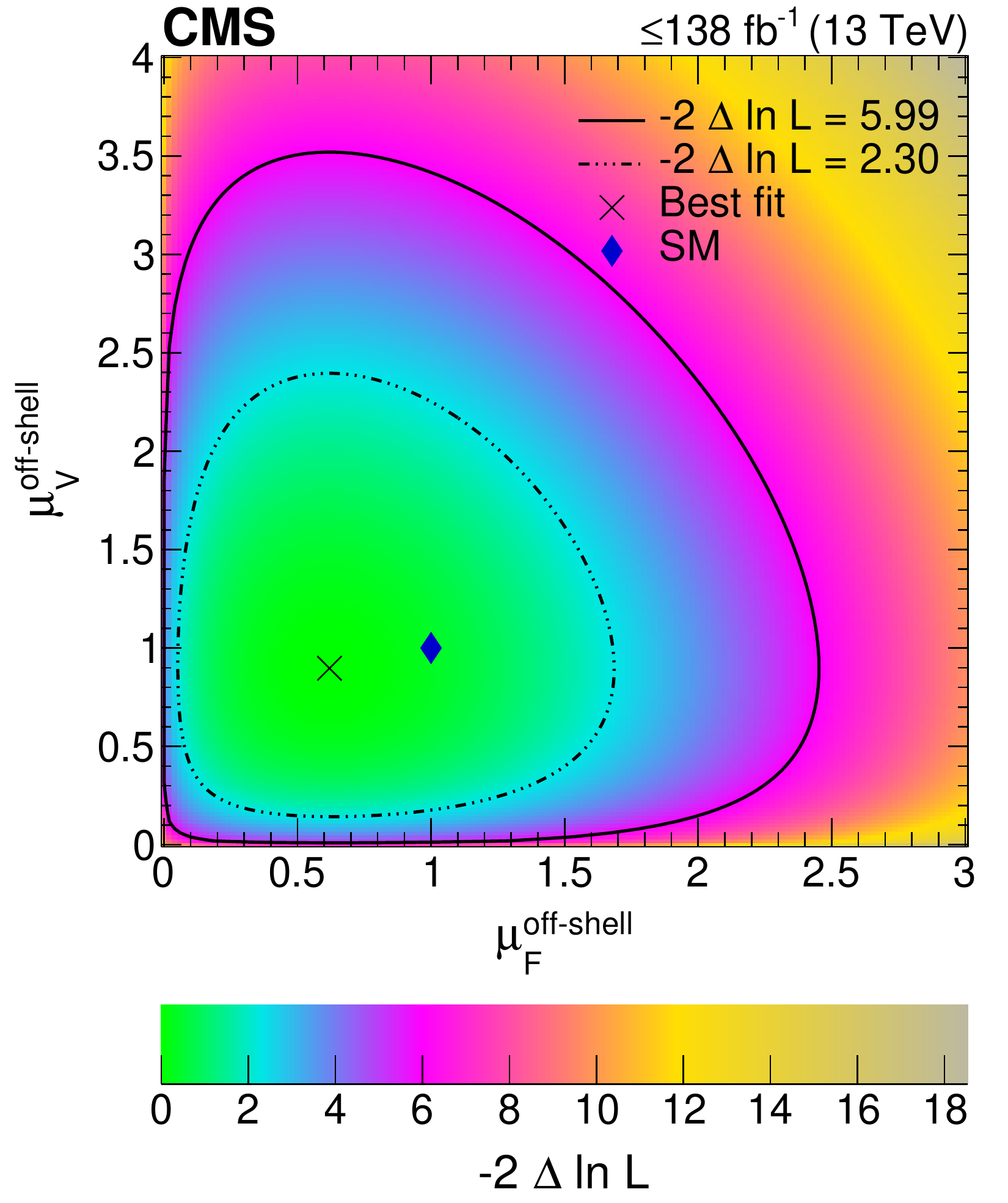}
\includegraphics[width=0.45\textwidth]{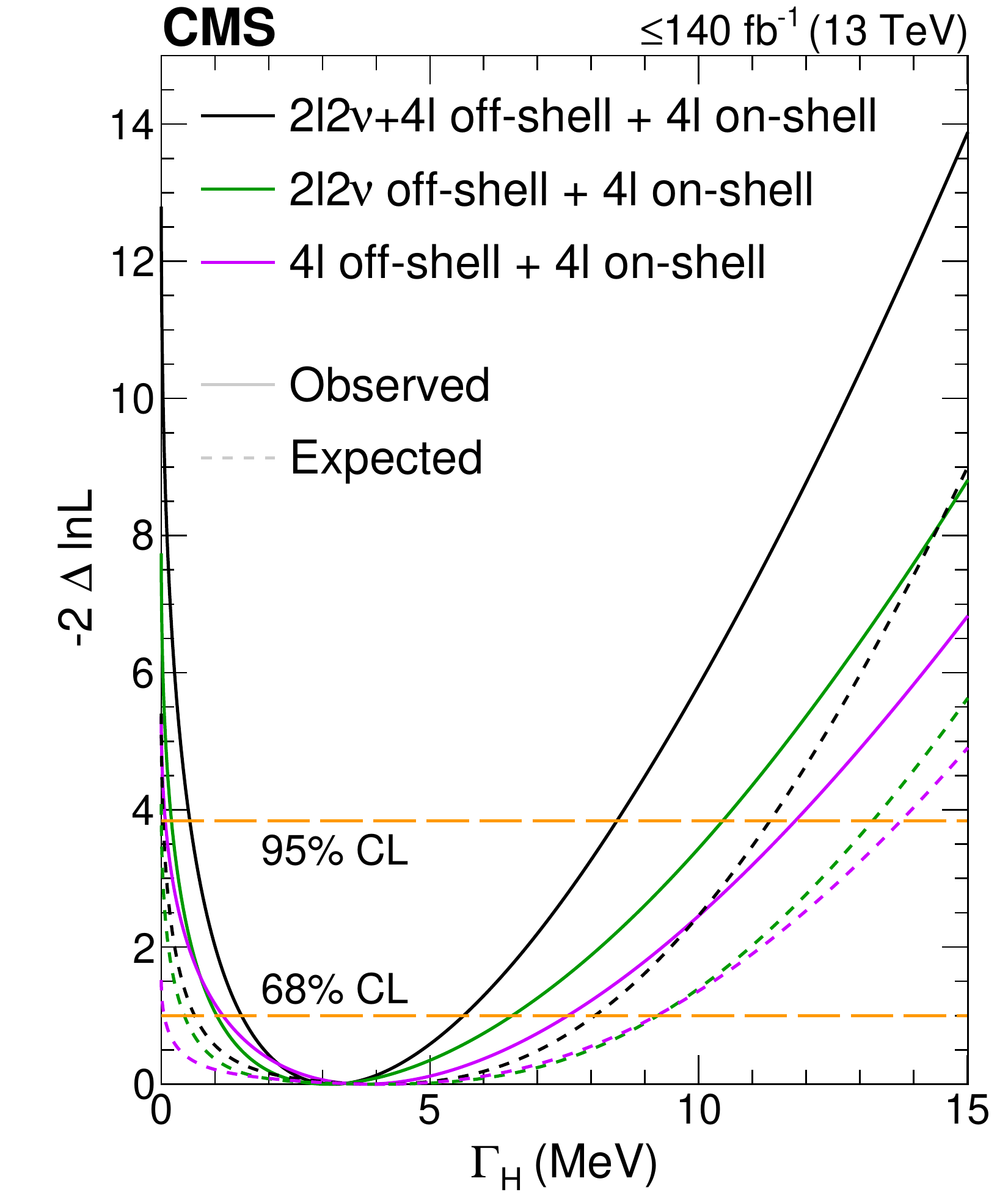}
\caption{
\textbf{Log-likelihood scans of $\muFoffsh$ and $\muVoffsh$, and $\GH$.}
Left panel:
Two-parameter likelihood scan of the \offshell \glufu and EW production signal strength parameters, $\muFoffsh$ and $\muVoffsh$, respectively.
The dot-dashed and dashed contours enclose the
68\% ($\dNLL=2.30$) and 95\% ($\dNLL=5.99$)~\CL regions.
The cross marks the minimum, and the blue diamond marks the SM expectation.
The integrated luminosity reaches only up to $\lumiDEF$ as \onshell $4\ell$ events are not included in performing this scan.
Right panel:
The observed (solid) and expected (dashed) one-parameter likelihood scans over $\GH$.
Scans are shown for the combination of $4\ell$ \onshell data with $4\ell$ \offshell (magenta) or $2\ell2\nu$ \offshell data (green) alone, or with both data sets (black).
The horizontal lines indicate the 68\% ($\dNLL=1.0$) and 95\% ($\dNLL=3.84$)~\CL regions.
The integrated luminosity reaches up to $\lumiCDEF$ as \onshell $4\ell$ events are included in performing these scans.
The exclusion of the no \offshell hypothesis is consistent with 3.6 standard deviations on both panels.
}
\label{fig:interp-2D-GH}
\end{figure*}

The profile likelihood scans in the
\muFoffsh and \muVoffsh plane are shown on the
left panel of Fig.~\ref{fig:interp-2D-GH}; scans over the individual signal strengths
are in Extended Data Fig.~\ref{fig:supp:interp}.
Likelihood scans over \GH are displayed in the right panel of Fig.~\ref{fig:interp-2D-GH}.
These scans always include information from the $4\ell$ \onshell data, and
the three cases displayed correspond to adding the $4\ell$ \offshell data alone, the $2\ell2\nu$ \offshell data alone, or adding both.
The steepness of the slope of the log-likelihood curves near $\muoffsh=0$ and $\GH=0\MeV$ is caused by 
the interference terms between the \Hboson and continuum \ZZ production amplitudes that scale with $\sqrt{\smash[b]{\muoffsh}}$ or $\sqrt{\GH{}}$, respectively.

The no~\offshell scenario with $\muoffsh=0$, or $\GH=0\MeV$ is excluded at a $p$-value of 0.0003 (3.6 standard deviations).
The $p$-value calculation is checked with pseudoexperiments and
the Feldman-Cousins prescription~\cite{Feldman:1997qc}.
As described in greater detail in the Methods section, the exclusion is illustrated in Extended Data Fig.~\ref{fig:supp:HypoLL} through a comparison of the total number of events in each \offshell signal region bin predicted for the fit of the data to the no~\offshell scenario, and the best fit.
Constraints on \GH are stable within 1\MeV (0.1\MeV{}) for the upper (lower) limits when testing the presence of anomalous \HVV couplings.
More results on these anomalous couplings are discussed in the Methods section, and can be found in Extended Data Fig.~\ref{fig:supp:interp} and Extended Data Table~\ref{table:supp:interp}. All results are also tabulated in the HEPData record for this analysis~\cite{hepdata}.

\begin{table}[!htb]
\centering
\topcaption
{
\textbf{Results on the \offshell signal strengths and $\GH$.}
The various fit conditions are indicated in the column labeled ``Cond.'':
Results on \muoffsh are presented with $\RVFoffsh=\muVoffsh / \muFoffsh$ either
unconstrained (u) or $=1$, and
constraints on $\muFoffsh$ and $\muVoffsh$ are shown with the other
signal strength unconstrained.
Results on \GH (in units of {\MeV{}}) are obtained with the \onshell signal strengths unconstrained, and the
different conditions listed for this quantity reflect which \offshell final states are combined with \onshell $4\ell$ data.
The expected central values, not quoted explicitly in this table, are either unity for \muoffsh{}, \muFoffsh{}, and \muVoffsh{}, or $\GH=4.1\MeV$.
}
\renewcommand{\arraystretch}{1.25}
\cmsTable{
\begin{tabular}{ccclll}
\multirow{2}{*}{Param.} & \multirow{2}{*}{Cond.} & \multicolumn{2}{c}{Observed} & \multicolumn{2}{c}{Expected}  \\
 & & $68\%$~\CL & \multicolumn{1}{c}{$95\%$~\CL} & \multicolumn{1}{c}{$68\%$~\CL} & \multicolumn{1}{c}{$95\%$~\CL}  \\
\hline
\muFoffshshort & $\muVoffshshort$ (u) & $0.62_{-0.45}^{+0.68}$ & $_{-0.614}^{+1.38}$ & $_{-0.99998}^{+1.1}$ & $\,<3.0$  \\
\muVoffshshort & $\muFoffshshort$ (u) & $0.90_{-0.59}^{+0.9}$ & $_{-0.849}^{+2.0}$ & $_{-0.89}^{+2.0}$ & $\,<4.5$  \\
\multirow{2}{*}{\muoffshshort}
                       & $\RVFoffshshort=1$ & $0.74_{-0.38}^{+0.56}$ & $_{-0.61}^{+1.06}$ & $_{-0.84}^{+1.0}$ & $_{-0.9914}^{+1.7}$  \\
                      & $\RVFoffshshort$ (u) & $0.62_{-0.45}^{+0.68}$ & $_{-0.6139}^{+1.38}$ & $_{-0.99996}^{+1.1}$ & $_{-0.99999}^{+2.0}$  \\
\GH             & $2\ell2\nu+4\ell$ & $3.2_{-1.7}^{+2.4}$ & $_{-2.7}^{+5.3}$ & $_{-3.5}^{+4.0}$ & $_{-4.07}^{+7.2}$ \\
\GH             & $2\ell2\nu$         & $3.1_{-2.1}^{+3.4}$ & $_{-2.9}^{+7.3}$ & $_{-3.7}^{+5.1}$ & $_{-4.099}^{+9.1}$ \\
\GH             & $4\ell$               & $3.8_{-2.7}^{+3.8}$ & $_{-3.73}^{+8.0}$ & $_{-4.05}^{+5.1}$ & $\,<13.8$ \\
\end{tabular}
}
\label{table:interp}
\end{table}

\clearpage

\bibliography{auto_generated}

\clearpage

{

\beginsupplement

\ifthenelse{\boolean{cms@external}}{\backmatter}{\fakesection{Methods}}
\section*{Methods}

\subsection*{Experimental setup}

The CMS apparatus~\cite{Chatrchyan:2008zzk} is a multipurpose, nearly hermetic detector, designed to trigger on~\cite{CMS:2016ngn} and identify muons, electrons, photons, and charged or neutral hadrons~\cite{Sirunyan:2018fpa,CMS:2020uim,Chatrchyan:2014pgm}. A global reconstruction algorithm, particle-flow (PF)~\cite{Sirunyan:2017ulk}, combines the information provided by the all-silicon inner tracker and by the crystal electromagnetic and brass-scintillator hadron calorimeters (ECAL and HCAL, respectively), operating inside a 3.8\unit{T} superconducting solenoid, with data from gas-ionization muon detectors interleaved with the solenoid return yoke, to build jets, missing transverse momentum, tau leptons, and other physics objects~\cite{Khachatryan:2016kdb,CMS:2019ctu,CMS:2018jrd}. In the following discussion up to likelihood scans, we will focus on the details of the $2\ell2\nu$ analysis. Analysis details for the \offshell $4\ell$ data can be found in Ref.~\cite{Sirunyan:2019twz}, 2015 \onshell $4\ell$ data in Refs.~\cite{Sirunyan:2017tqd,Sirunyan:2019twz}, and 2016--2018 \onshell $4\ell$ data in Ref.~\cite{Sirunyan:2021fpv}.

\subsection*{Physics objects}

Events in the $2\ell2\nu$ signal region, the \emu CR, and the
trilepton \WZ CR are selected
using single-lepton and dilepton triggers. The efficiencies of 
these selections are measured using orthogonal triggers, \ie, jet or \ptmiss triggers,
and events triggered on a third, isolated lepton, or a jet. They range between 78\% and
100\%, depending on the flavor of the leptons, and $\pt$ and $\eta$ of the dilepton system, taking lower values at lower $\pt$.
Photon triggers are used to collect events for the \Gamjetstxt CR.
The photon trigger efficiency is measured using a tag-and-probe method~\cite{CMS:2010svw}
in $\PZ\to\cPe\cPe$ events, with one electron interpreted as a photon with tracks ignored,
as well as through a study of \llGam events.
The efficiency is found to range from ${\sim}55\%$ at $55\GeV$ in photon $\pt$ to
${\sim}95\%$ at photon $\pt>220\GeV$.

Jets are reconstructed using the anti-$\kt$ algorithm~\cite{Cacciari:2011ma} with
a distance parameter of 0.4.
Jet energies are corrected for instrumental effects, as well as for the contribution of particles originating from
additional $\Pp{}\Pp{}$ interactions (\pileup).
A multivariate technique is used to suppress jets from \pileup
interactions~\cite{Sirunyan:2020foa}.
For the purpose of this analysis, we select jets of $\pt > 30\GeV$ and $\abs{\eta}<4.7$, and they must be
separated by $\Delta R=\sqrt{\smash[b]{(\Delta \phi)^2 + (\Delta \eta)^2}}>0.4$, with $\phi$ being the azimuthal angle measured in radians, from a lepton or a photon of interest.
Jets within $\abs{\eta}<2.5$ ($\abs{\eta}<2.4$ for 2016 data) can be
identified as \PQb jets
using the $\textsc{DeepJet}$ algorithm~\cite{Bols:2020bkb} with a loose working point.
The efficiency of this working point ranges between $75\%$ and $95\%$,
depending on $\pt$, $\eta$, and the data period.

The missing transverse momentum vector \ptvecmiss is estimated
from the negative of the vector sum of the transverse momenta of all
PF candidates.
Dedicated algorithms~\cite{Chatrchyan:2011tn} are used to eliminate events featuring
cosmic ray contributions, beam-gas interactions, beam halo, or calorimetric noise.

The algorithms to reconstruct leptons are described in detail in Ref.~\cite{Sirunyan:2018fpa} for muons and Ref.~\cite{CMS:2020uim} for electrons.
Muons are identified using a set of requirements on individual variables,
while electrons are identified using a boosted decision tree algorithm.
Leptons of interest in this analysis are expected to be isolated with
respect to the activity in the rest of the event.  A measure of isolation  
is computed from the flux of photons and hadrons reconstructed by the PF
algorithm that are
within a cone of $\Delta R<0.3$ built around the lepton
direction, including corrections from the contributions of \pileup.
We define loose and tight isolation requirements for muons (electrons)
with $\pt > 5\GeV$ and $\abs{\eta}<2.4$ ($\abs{\eta}<2.5$).
The efficiency of loose selection for muons (electrons) ranges from
${\sim}85\%$ (65--75\%, depending on $\eta$) at $\pt=5\GeV$ to $>90\%$ ($>85\%$) at $\pt>25\GeV$.
The additional requirements for tight selections reduce efficiencies by 10--15\%.

Photons are reconstructed from energy clusters in the ECAL not linked
to charged tracks, with the exception of converted
photons~\cite{CMS:2020uim}.
Their energies
are corrected for shower containment in the ECAL crystals
and energy loss due to conversions in the tracker with a multivariate
regression.
In this analysis, we consider
photons with $\pt > 20\GeV$ and $\abs{\eta}$ up to 2.5, with
requirements on shower shape and isolation used to identify isolated photons and separate them from hadronic jets.
The selection requirements are tightened in the \Gamjetstxt CR, which leads to selection efficiencies in the range 50--75\%, depending on $\pt$ and $\eta$.

\subsection*{Event simulation}

{\tolerance=600
The signal Monte Carlo (MC) samples are generated for an undecayed \Hboson for
\glufu{}, \VBF{}, \ZH{}, and \WH
productions using the \POWHEG~2~\cite{Frixione:2007vw,Nason:2009ai,Bagnaschi:2011tu,Luisoni:2013cuh}
program
at next-to-leading order (NLO) in QCD at various \Hboson pole masses, ranging from $125\GeV$ to $3\TeV$.
The generated \PH bosons are decayed to
four-fermion final states through intermediate
\PZ bosons using the \jhugen~\cite{Gritsan:2020pib} program, with versions between 6.9.8 and 7.4.0.
\par}

These samples are reweighted using the \mela matrix element package, which interfaces with the \jhugen and \MCFM~\cite{MCFM,Campbell:2011bn,Campbell:2013una,Campbell:2015vwa} matrix elements, following the same reweighting techniques used in
Ref.~\cite{Sirunyan:2019twz} to obtain the final \ZZ event sample, including the \Hboson
contribution, the continuum, and their interference.
The \MelaAnalytics package developed for Ref.~\cite{Sirunyan:2019twz} is used to automate
matrix element computations and to account for the extra partons in the NLO simulation.
The \glufu generation is rescaled with the next-to-NLO (NNLO) QCD K-factor, differential in $\mVV$,
and an additional uniform K-factor of 1.10 for the next-to-NNLO cross section computed at $\mH=125\GeV$~\cite{deFlorian:2016spz}. Furthermore, the pole mass values of the top quark ($173\GeV$) and the bottom quark ($4.8\GeV$)~\cite{Zyla:2020zbs} are used in the massive loop calculations for the generation of this process. The difference that would be introduced by using the $\overline{\mathrm{MS}}$ renormalization scheme for these masses is found to be within the systematic uncertainties after accounting for the effects on both the \Hboson and continuum \ZZ amplitudes.

The tree-level Feynman diagrams in
Fig.~\ref{fig:feyngg}
illustrate
the complete set contributing to the $\glufu \to \ZZ$ process
on the leftmost top and bottom panels, and
some of the diagrams contributing to
the EW $\ZZ$ production associated with two fermions on the middle and top right panels.
Extended Data Figs.~\ref{fig:supp:feyn-EWsig} and \ref{fig:supp:feyn-EWcontin}
display the full set of diagrams for the EW process.

The $\qqbar \to \ZZ$ and $\WZ$ MC samples are also generated with \POWHEG~2 applying
EW NLO corrections for two \onshell $\PZ$ and $\PW$ bosons~\cite{Bierweiler:2013dja,Manohar:2016nzj},
and NNLO QCD corrections as a function of
$\mVV$~\cite{Grazzini:2017mhc}.
The tree-level Feynman diagrams for these noninterfering continuum contributions are illustrated in Extended Data Fig.~\ref{fig:supp:feyn-qqVZ}.
Samples for the \tZXtxt processes, or other processes contributing to the CRs,
are generated using \MGvATNLO at NLO or LO precision using the FxFx~\cite{Frederix:2012ps} or MLM~\cite{Alwall:2007fs} schemes, respectively, to match jets from matrix element calculations and parton shower.

The parton shower and hadronization are modeled with \PYTHIA (8.205 or 8.230)~\cite{Sjostrand:2014zea},
using tunes \textsc{CUETP8M1}~\cite{Khachatryan:2015pea} for the 2015 and 2016 data sets, and \textsc{CP5}~\cite{Sirunyan:2019dfx} for the 2017 and 2018 periods.
The PDFs are taken from NNPDF 3.0~\cite{Ball:2014uwa} 
with QCD orders matching those of the
cross section calculations.
Finally, the detector response is simulated with the
\GEANTfour~\cite{Agostinelli2003250} package.

\subsection*{Signal region selection requirements}

Events in the $2\ell2\nu$ final state are
required to have two opposite-sign, same-flavor leptons ($\PGmp\PGmm$ or $\Pep\Pem$)
satisfying tight isolation requirements
with $\pt > 25\GeV$, \mll within $15\GeV$ of $\mZ$,
and $\ptll > 55\GeV$.
Additional requirements are imposed to reduce contributions
from \Zjetstxt and \ttbar processes as follows.
Events with $\PQb$-tagged jets, additional loosely isolated
leptons of $\pt > 5\GeV$, or additional loosely identified photons with $\pt > 20\GeV$ are vetoed.
To further improve the effectiveness of the lepton
veto, events with isolated reconstructed tracks
of $\pt > 10\GeV$ are removed.
This requirement is also effective against one-prong $\tau$ decays.

The value of \ptmiss is required to be $>125\GeV$ ($>140\GeV$) for $\Nj < 2$
($\geq 2$).
Requirements are imposed on the unsigned azimuthal opening
angles ($\Delta\phi$) between \ptvecmiss and other objects in the event
in order to reduce contamination from \ptmiss misreconstruction:
$\dphillmet > 1.0$ between \ptvecmiss and \vecptll,
$\dphilljetsmet > 2.5$ between \ptvecmiss and $\vecptll + \sum \vecptj$,
$\mindphijetmet > 0.25$ (0.50) between \ptvecmiss and \vecptj for $\Nj = 1$
($\Nj \geq 2$), where \vecptj is the transverse momentum vector of a jet.

Finally, events are split into lepton flavor ($\PGm\PGm$ or $\Pe\Pe$)
and jet multiplicity ($\Nj = 0, 1, \geq 2$) categories.
The resulting event distributions are illustrated along the $\mTZZ$ observable
in Extended Data Fig.~\ref{fig:supp:mTZZ-SR}.

\subsection*{Matrix element kinematic discriminants}

In events with $\Nj \geq 2$,
we use two \mela kinematic discriminants for the \VBF process,
$\DjjVBF$
and
$\DjjVBFHS$~\cite{Sirunyan:2019twz}.
Each of these discriminants consists of a ratio of two matrix elements,
or equivalently a ratio of event-by-event probability functions, expressed in terms of the
four-momenta of the \Hboson and the two jets leading in $\pt$.
The four-momentum of the \Hboson in the $2\ell2\nu$ channel
is approximated by taking
the $\eta$
of the $\PZ \to 2\nu$ candidate, together with its sign, to be the same as that of the
$\PZ \to 2\ell$ candidate. This approximation is found to be adequate through MC studies.

In both discriminants, one of matrix elements is always computed for
the SM \Hboson production through gluon fusion.
The remaining matrix element is computed for the SM
\VBF process in $\DjjVBF$,
so this discriminant improves the sensitivity to the EW \Hboson production.
The $\DjjVBFHS$ discriminant also computes the remaining matrix element for the \VBF process,
but under the $\AC{2}$ \HVV coupling hypothesis instead of the SM scenario.
We find that this second discriminant brings additional sensitivity to SM backgrounds
as well as being sensitive to the
$\AC{2}$ \HVV coupling hypothesis by design.
When anomalous \HVV contributions are considered,
the $\AC{2}$ hypothesis used in the computation is replaced by the
appropriate $\ai$ hypothesis to optimize sensitivity for the coupling of interest.

\subsection*{Control regions}

As already mentioned, 
\Zjetstxt events are a background to the
$2\ell2\nu$ signal selection.  This can occur because of
resolution effects in \ptmiss and the large cross section for this process.  Since \Gamjetstxt and \Zjetstxt
have similar production and \ptmiss resolution properties,
the \Zjetstxt contributions at high
\ptmiss can be estimated from a \Gamjetstxt
CR~\cite{Chatrchyan:2012iqa}.

In this CR, all event selection requirements are the same as those on the
signal region, except that the photon replaces the
$\PZ \to \ell \ell$ decay.
The \mTZZ kinematic variable is constructed using
the photon $\pt$ in place of $\ptll$, and \mZ in place of \mll.
Only photons in the barrel region
(\ie, $\abs{\eta}<1.44$) are considered for
$\Nj < 2$ to eliminate beam halo events that can mimic the
$\gamma + \ptmiss$ signature.
Reweighting factors are extracted as a function of 
photon \pt, 
photon $\eta$ (when $\Nj\ge 2$), and
the number of observed $\Pp{}\Pp{}$ collisions by matching the corresponding
distributions in \Gamjetstxt sidebands at low $\ptmiss$ ($< 125\GeV$) to those of \Zjetstxt sidebands with the same requirement at each $\Nj$ category separately.  These
reweighting factors are then applied to the high-$\ptmiss$ \Gamjetstxt data sample.
This technique to estimate the background from the data is verified using closure tests from the simulation by comparing the \Zjetstxt and reweighted \Gamjetstxt MC distributions over each kinematic observable.

Contributions to the \Gamjetstxt CR from events with genuine,
large \ptmiss from the \ZnunuGam{}, \WlnuGam{}, and \Wlnujetstxt processes are
subtracted in the final estimate of the instrumental \ptmiss background.
The first two are estimated from simulation,
where the \ZGam contribution is corrected based on the
observed rate of \ZllGam{}.
The \Wjetstxt contribution is estimated from a single-electron
sample selected with requirements similar to those in the \Gamjetstxt CR.
Representative distributions for this estimate are shown in Extended Data Fig.~\ref{fig:supp:InstrMET}.

Processes such as $\Pp\Pp \to \ttbar$ and
$\Pp\Pp \to \WW$, including nonresonant \Hboson contributions, can produce
two leptons and large $\ptmiss$ without a resonant $\PZ \to \ell\ell$ decay.
The kinematic properties of the dilepton system in these processes is the same
for any combination of lepton flavors $\Pe$ or $\PGm$.
These nonresonant $\Pe\Pe$ or $\PGm\PGm$ background processes are therefore estimated from an \emu CR.
This CR is constructed applying the same requirements used in the signal
selection except for the flavor of the leptons.
Data events are reweighted
to account for differences in trigger and
reconstruction efficiencies between $\Pe\PGm$, and $\Pe\Pe$ or $\PGm\PGm$ final states.
Representative distributions for this estimate are shown in Extended Data Fig.~\ref{fig:supp:NRB}.

A third CR selects trilepton $\qqbar \to \WZ$ events.
These events are used to constrain the
normalization and kinematic properties of the
$\qqbar \to \ZZ$ and $\WZ$ continuum contributions.
The $\PZ \to \ell\ell$ candidate is identified from the opposite-sign, same-flavor lepton pair with $\mll$ closest to $\mZ$, and
the value of $\mll$ for this $\PZ$ candidate is required to be within $15\GeV$ of $\mZ$.
Trigger requirements are only placed on this $\PZ$ candidate.
The remaining lepton is identified as the lepton from the $\PW$ decay ($\ell_{\PW{}}$).
The leading-$\pt$ lepton from the $\PZ$ decay is required to satisfy $\pt > 30\GeV$, and the remaining leptons are required to satisfy $\pt > 20\GeV$.

Similar to the signal region,
requirements are imposed on the unsigned $\Delta\phi$ between \ptvecmiss and other objects in the event
in order to reduce contamination from the \Zjetstxt and $\qqbar \to \PZ \gamma$ processes:
$\dphillmet > 1.0$ between \ptvecmiss and $\vecptll$ for the $\PZ$ candidate,
$\dphiWZjetsmet > 2.5$ between \ptvecmiss and $\vec{p}_{\mathrm{T}}^{\,3\ell} + \sum \vecptj$, and
$\mindphijetmet > 0.25$ between \ptvecmiss and $\vecptj$.

The $\PW$ boson transverse mass is defined through the vector transverse momentum of $\ell_{\PW{}}$, $\vec{p}_{\mathrm{T}}^{\ell_\PW}$, as $\mT^{\ell_{\PW{}}}=\sqrt{\smash[b]{2 ( p_{\mathrm{T}}^{\ell_{\PW}} \ptmiss - {\vec p}_{\mathrm{T}}^{\ell_\PW} \cdot \ptvecmiss )}}$, and additional requirements are imposed on $\ptmiss$ and $\mT^{\ell_{\PW{}}}$ in order to reduce contamination from the \Zjetstxt and $\qqbar \to \PZ \gamma$ processes further: $\ptmiss>20\GeV$, $\mT^{\ell_{\PW{}}}>20\GeV$ ($10\GeV$) for $\ell_{\PW}=\cPm$ ($\cPe$), and $\mathrm{A} \times \mT^{\ell_{\PW}} + \ptmiss > 120\GeV$, with $\mathrm{A}=1.6$ ($4/3$) for $\ell_{\PW}=\cPm$ ($\cPe$).
All other requirements on $\PQb$-tagged jets, and additional leptons or photons are the same as those for the signal region.

The events are finally split into categories of the flavor of $\ell_{\PW}$ ($\cPm$ or $\cPe$) and jet multiplicity ($\Nj = 0, 1, \geq 2$), and binned in $\mTWZ$, defined using the $\PW$ boson mass $\mW=80.4\GeV$~\cite{Zyla:2020zbs} as
\ifthenelse{\boolean{cms@external}}{
\begin{linenomath}
\begin{equation*}
\begin{aligned}
  {\left( \mTWZ \right)}^2 &=
  \Biggl[ \sqrt{{\ptll}^2+{\mll}^2} \\
  &+ \sqrt{\smash[b]{{ \big\lvert \ptvecmiss+\vec{p}_{\mathrm{T}}^{\ell_\PW} \big\rvert}^2+{\mW}^2}} \Biggr]^2 \\
  &- {\bigg\lvert \vecptll + \ptvecmiss + \vec{p}_{\mathrm{T}}^{\ell_\PW} \bigg\rvert}^2.  
\end{aligned}
\label{eq:mTWZ}
\end{equation*}
\end{linenomath}
}{
\begin{equation*}
 {\left( \mTWZ \right)}^2 =
 \left[ \sqrt{{\ptll}^2+{\mll}^2}
 + \sqrt{{ \left| \ptvecmiss+\vec{p}_{\mathrm{T}}^{\ell_\PW} \right|}^2+{\mW}^2} \right]^2
 - {\left| \vecptll + \ptvecmiss + \vec{p}_{\mathrm{T}}^{\ell_\PW} \right|}^2.
\label{eq:mTWZ}
\end{equation*}
}
Event distributions along $\mTWZ$ from this CR are shown in Extended Data Fig.~\ref{fig:supp:WZCR}.

\subsection*{Likelihood scans}

As mentioned in the discussion of data interpretation, the likelihood is constructed from several multidimensional distributions binned over the different event categories.
Profile likelihood scans over \muFoffsh{}, \muVoffsh{}, \muoffsh{}, and \GH
are shown in Extended Data Fig.~\ref{fig:supp:interp}.
When testing the effects of anomalous \HVV couplings, we perform fits to the data
with all BSM couplings set to zero, except the one being tested, in the
model to be fit.
Because the only remaining degree of freedom is the ratio of these BSM couplings to the SM-like coupling, \AC{1},
the probability densities are
parametrized in terms of the effective, signed \onshell cross section fraction $\fcospai$ for each of the \ai coupling,
where the sign of the
phase of \ai relative to $\AC{1}$ is absorbed into the definition of $\fcospai$~\cite{Sirunyan:2021fpv}.
The constraints on \GH are found to be stable within 1\MeV (0.1\MeV{}) for the upper (lower) limits under the different anomalous \HVV coupling conditions, and they are summarized in Extended Data Table~\ref{table:supp:interp}.

In addition, we provide a simplified illustration for the exclusion of the no \offshell hypothesis in Extended Data Fig.~\ref{fig:supp:HypoLL}.
In this figure, the total number of events in each bin of the likelihood are compared
from the $2\ell2\nu$ and $4\ell$ \offshell regions
for the fit of the data to the no~\offshell ($N_{\text{no~\offshell}}$) scenario,
and the best fit ($N_{\text{best~fit}}$).
Events can then be rebinned over the ratio
$N_{\text{no \offshell}}/ ( N_{\text{no \offshell}}+N_{\text{best fit}})$ extracted from each bin,
and these rebinned distributions can then be compared at different $\GH$ values.
In particular, we compare the observed and expected event distributions over this ratio under the best fit scenario, and the scenario with no \offshell \Hboson production,
in order to illustrate which bins bring most sensitivity to the exclusion of the no \offshell scenario.
The exclusion is noted to be most apparent from the last two bins displayed in this figure. We note, however, that the full power of the analysis ultimately comes from the different bins over the multidimensional likelihood, and that this figure only serves to condense the information for illustration.

When we perform separate likelihood scans over
the three \fcospai fractions,
only the corresponding BSM parameter is allowed to be nonzero in the fit.
Profile likelihood scans for \fcospAC{2}, \fcospAC{3} and \fcospLC{1}
under different fit conditions are shown in Extended Data
Fig.~\ref{fig:supp:interp}, and
the summary of the allowed intervals at 68\% and 95\%~\CL is presented in Extended Data Table~\ref{table:supp:interp}.

\ifthenelse{\boolean{cms@external}}{\backmatter}{\fakesection{Data availability}}
\section*{Data availability}
Tabulated results are provided in the \href{https://dx.doi.org/10.17182/hepdata.127288}{HEPData record} for this analysis~\cite{hepdata}. Release and preservation of data used by the
CMS collaboration as the basis for publications is guided by the \href{https://cms-docdb.cern.ch/cgi-bin/PublicDocDB/RetrieveFile?docid=6032&filename=CMSDataPolicyV1.2.pdf&version=2}{CMS data preservation, reuse, and open acess policy}.

\ifthenelse{\boolean{cms@external}}{\backmatter}{\fakesection{Code availability}}
\section*{Code availability}
The CMS core software is publicly available on GitHub (\url{https://github.com/cms-sw/cmssw}).

\ifthenelse{\boolean{cms@external}}{
\clearpage
}{}

\begin{figure*}[htbp]
\centering
\includegraphics[width=.6\textwidth]{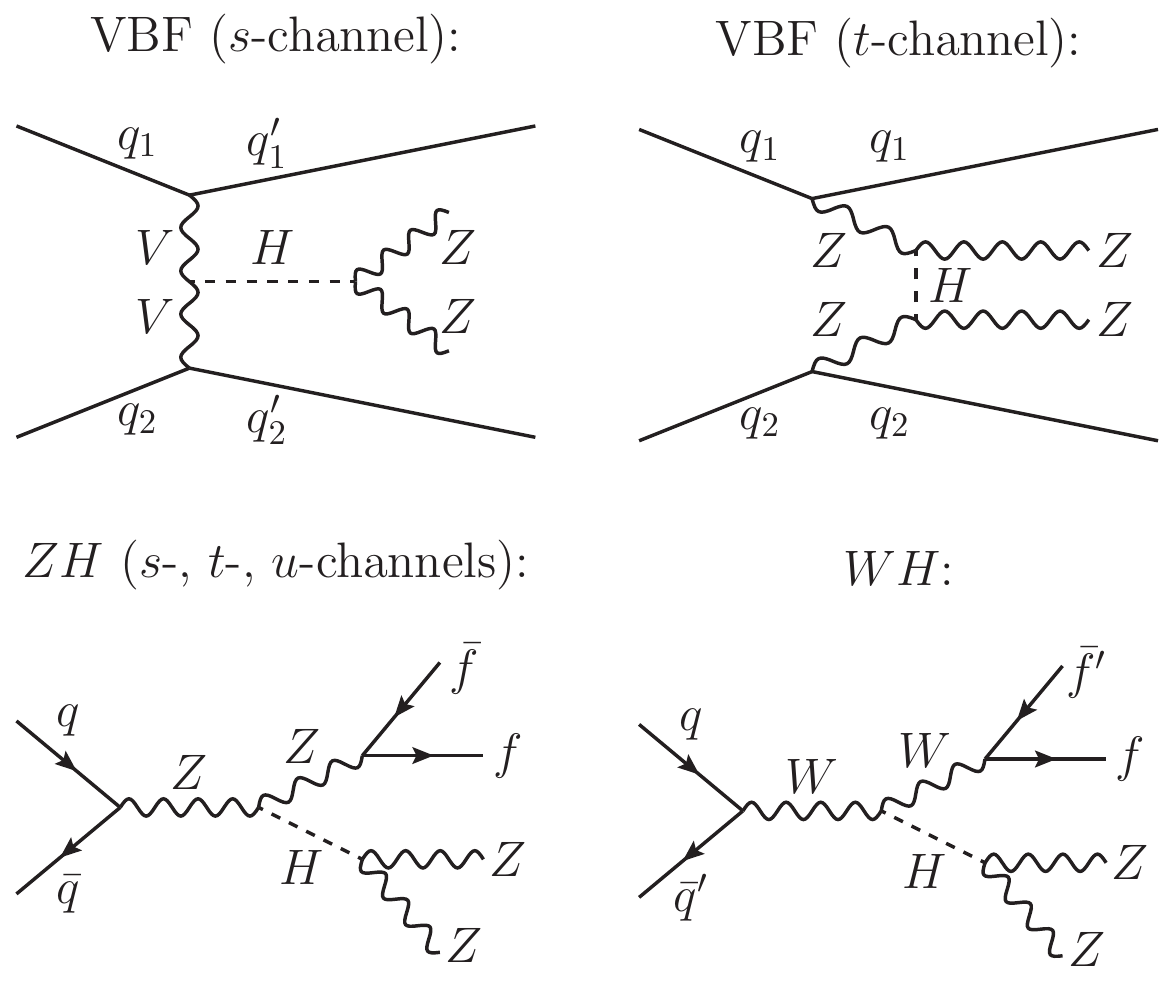}
\caption{
\textbf{Feynman diagrams for the \Hboson-mediated EW \ZZ production contributions.}
Here, $f$ refers to any $\ell$, $\nu$, or $q$. The tree-level diagrams featuring \VBF production are grouped together in the upper row, and those featuring $\VH$ production are grouped in the lower row.
The interaction displayed in each diagram is meant to progress from left to right. Each straight, curvy, or curly line refers to the different set of particles denoted. Straight, solid lines with no arrows indicate the line could refer to either a particle or an antiparticle, whereas those with forward (backward) arrows refer to a particle (an antiparticle).
}
\label{fig:supp:feyn-EWsig}
\end{figure*}

\ifthenelse{\boolean{cms@external}}{
\clearpage
}{}

\begin{figure*}[htbp]
\centering
\includegraphics[width=.6\textwidth]{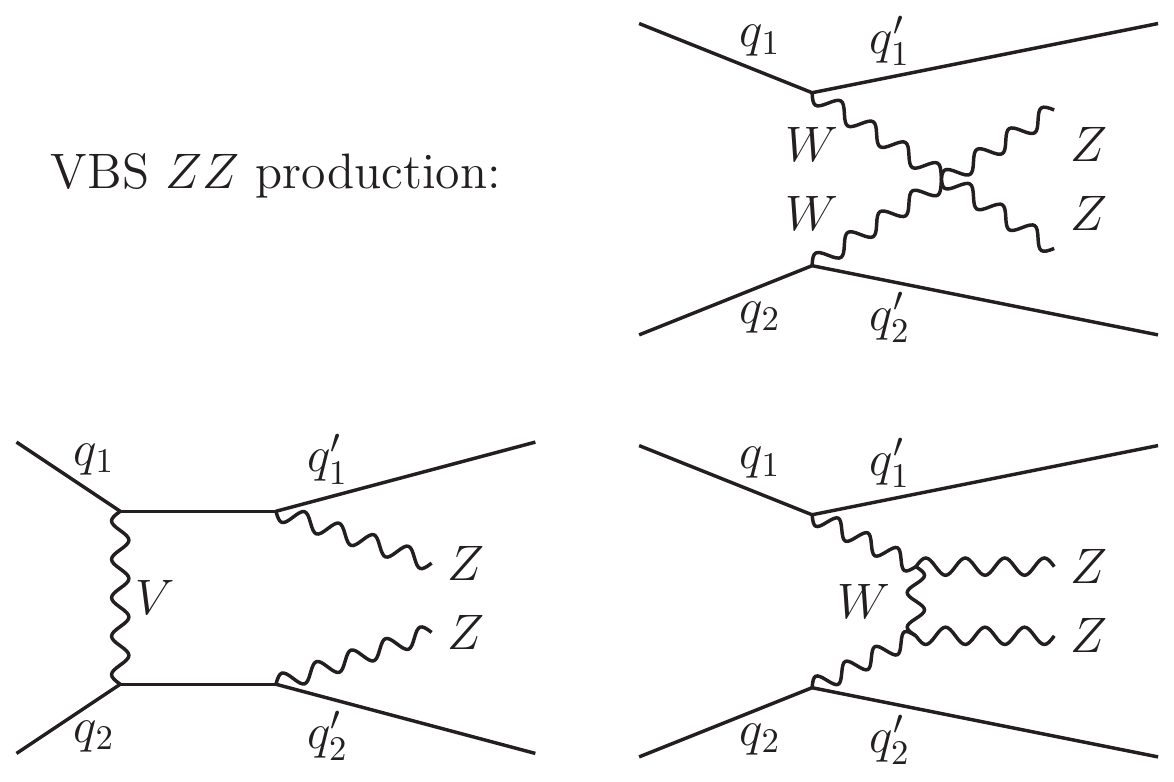} \\
\includegraphics[width=.6\textwidth]{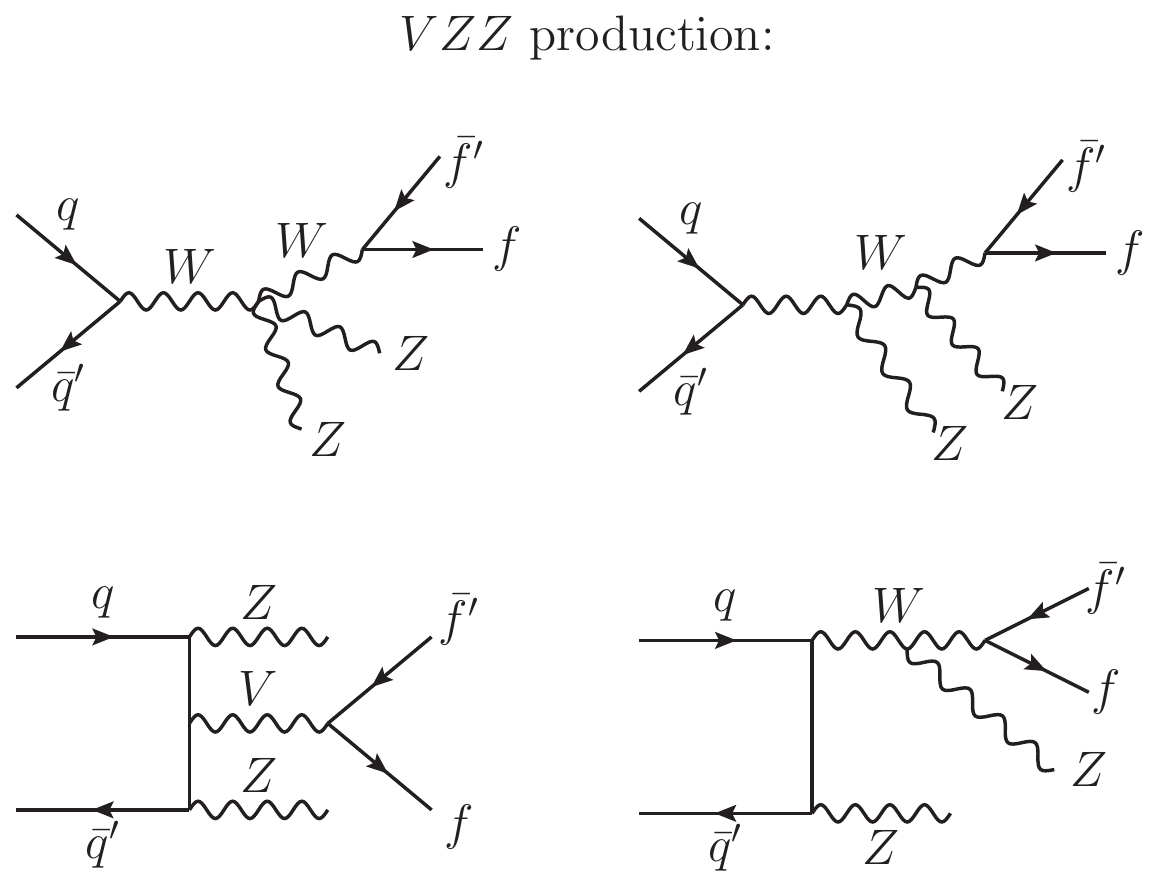}
\caption{
\textbf{Feynman diagrams for the EW continuum $\ZZ$ production contributions.}
Here, $f$ refers to any $\ell$, $\nu$, or $q$. The tree-level diagrams featuring vector boson scattering (VBS) production are grouped together in the upper half, and those featuring $\V{}\ZZ$ production are grouped in the lower half.
The interaction displayed in each diagram is meant to progress from left to right. Each straight, curvy, or curly line refers to the different set of particles denoted. Straight, solid lines with no arrows indicate the line could refer to either a particle or an antiparticle, whereas those with forward (backward) arrows refer to a particle (an antiparticle).
}
\label{fig:supp:feyn-EWcontin}
\end{figure*}

\ifthenelse{\boolean{cms@external}}{
\clearpage
}{}

\begin{figure*}[htbp]
\centering
\includegraphics[width=.18\textwidth]{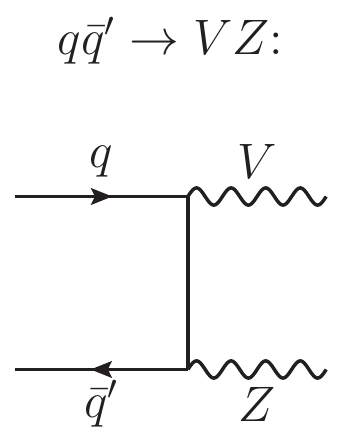}
\caption{
\textbf{Feynman diagram for the $\qqZZ$ and $\qqWZ$ processes.}
Both processes are represented at tree level with a single diagram. These two processes constitute the major irreducible, noninterfering background contributions in the \offshell region.
The interaction displayed in each diagram is meant to progress from left to right. Each straight, curvy, or curly line refers to the different set of particles denoted. Straight, solid lines with no arrows indicate the line could refer to either a particle or an antiparticle, whereas those with forward (backward) arrows refer to a particle (an antiparticle).
}
\label{fig:supp:feyn-qqVZ}
\end{figure*}

\ifthenelse{\boolean{cms@external}}{
\clearpage
}{}

\begin{figure*}[htbp]
\centering
\includegraphics[width=.3\textwidth]{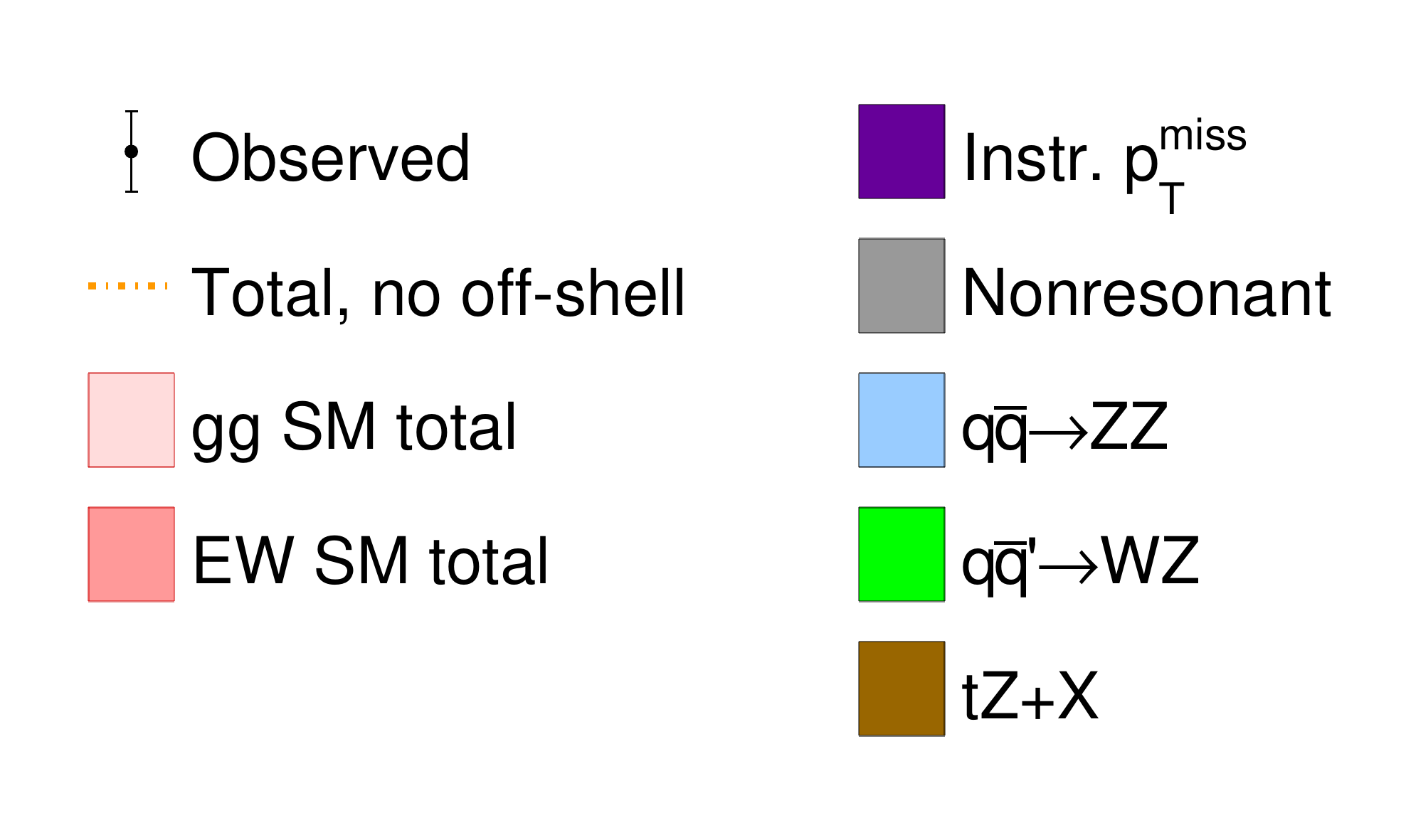} \\
\includegraphics[width=.3\textwidth]{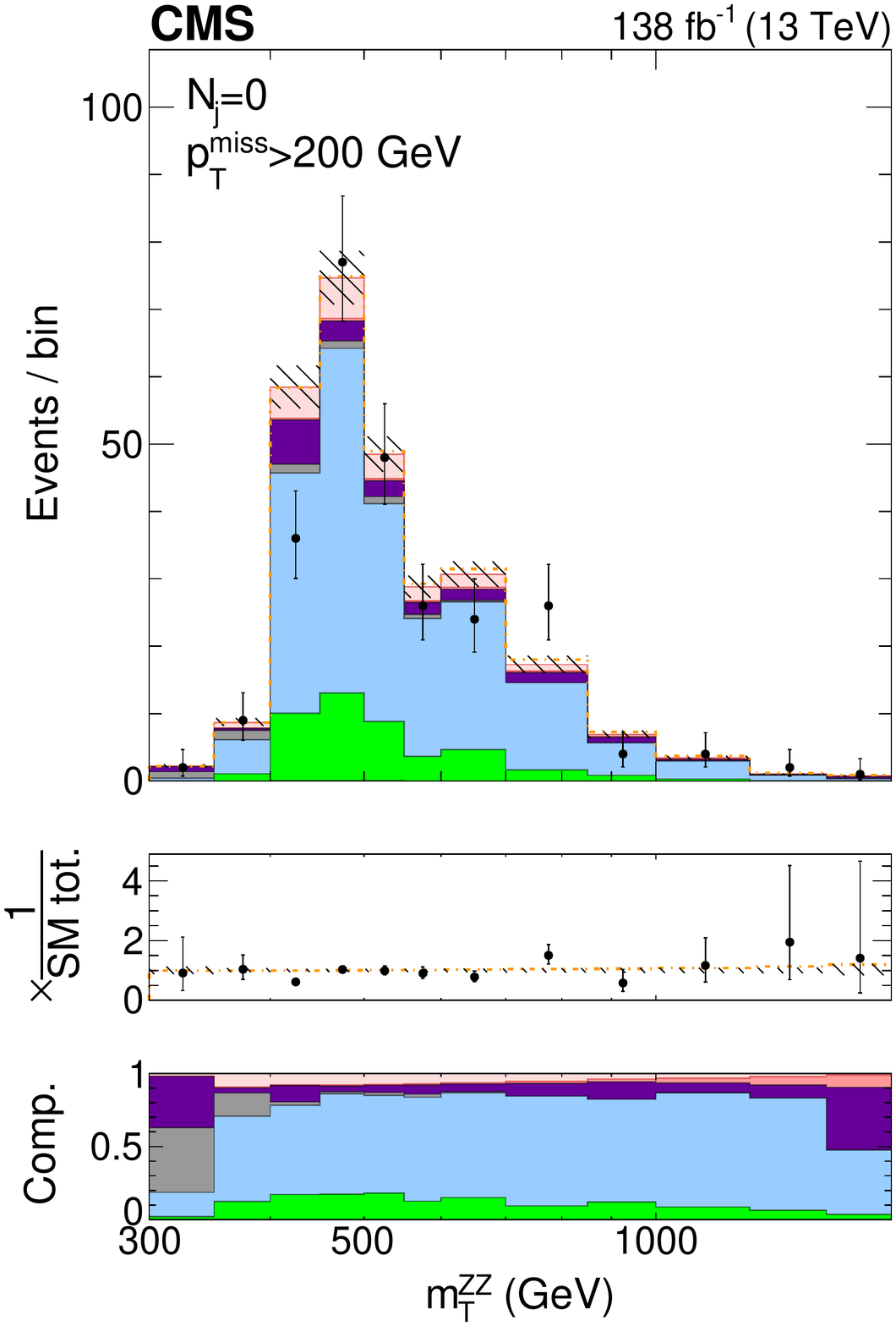}
\includegraphics[width=.3\textwidth]{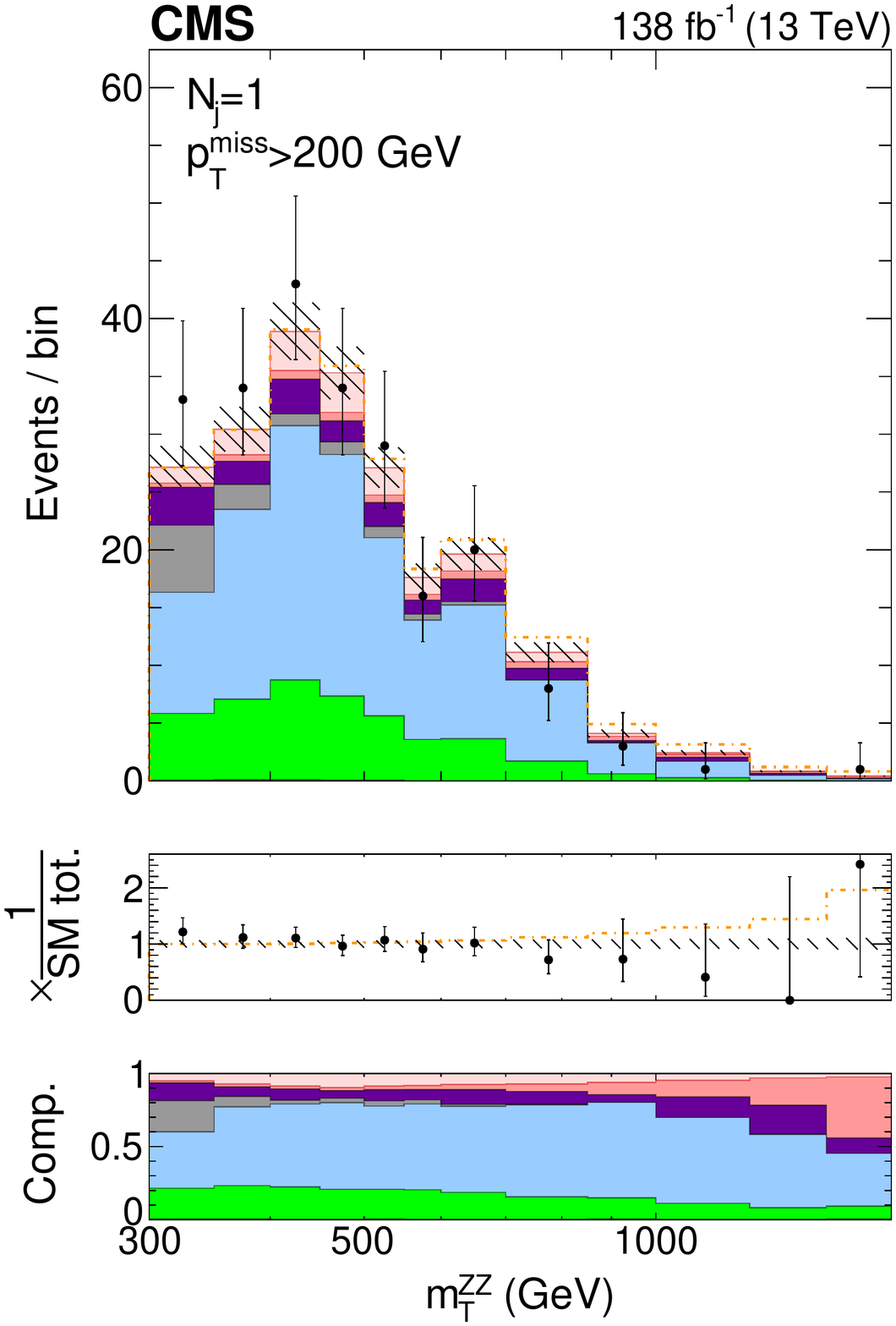}
\includegraphics[width=.3\textwidth]{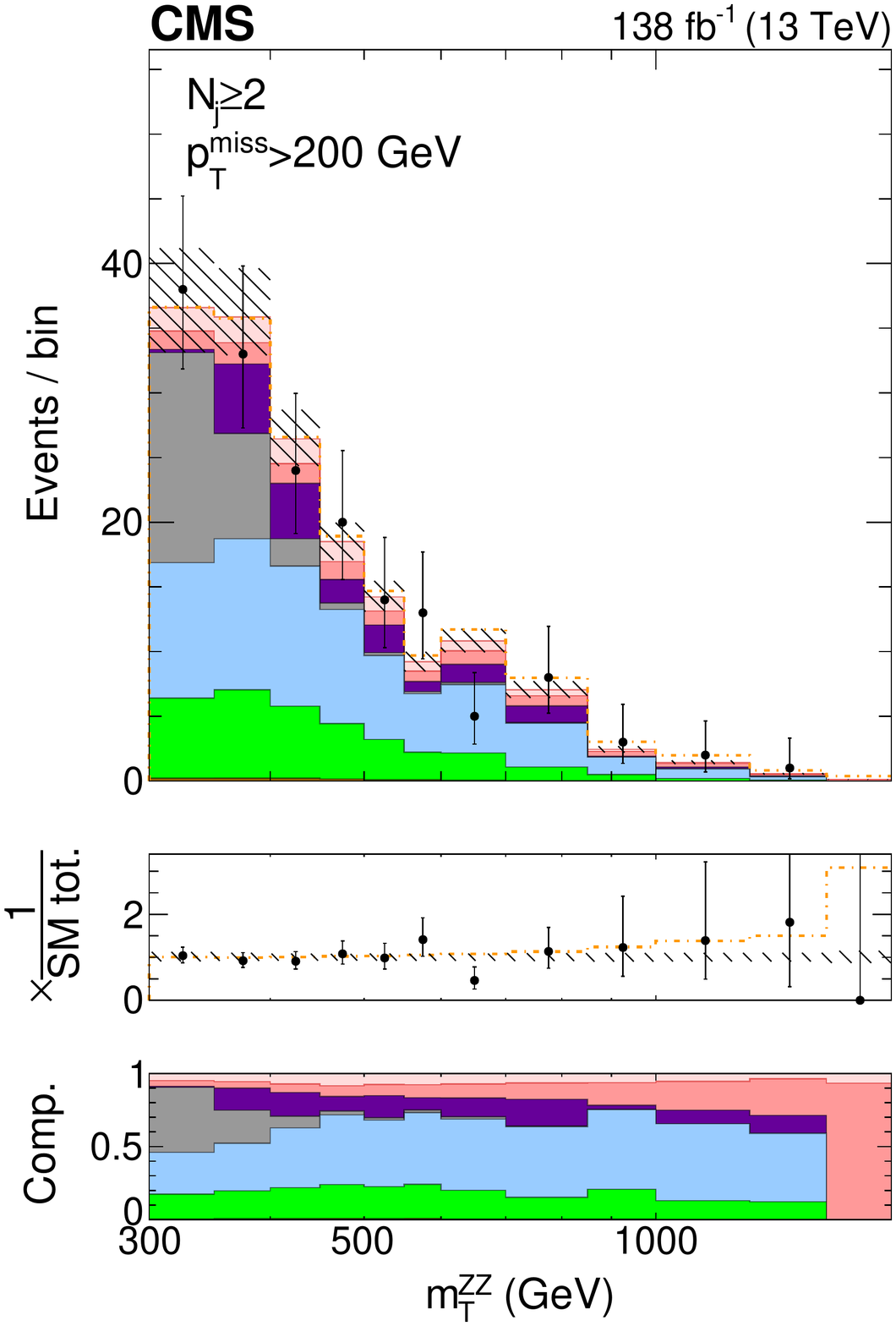}
\caption{
\textbf{Distributions of \mTZZ in the different $\Nj$ categories of the $2\ell2\nu$ signal region.}
The postfit distributions of the transverse \ZZ invariant mass are displayed in the jet multiplicity categories of $\Nj=0$ (left), $=1$ (middle), and $\geq 2$ (right) with a missing transverse momentum requirement of $\ptmiss>200\GeV$ to enrich \Hboson contributions.
The color legend for the stacked or dot-dashed histograms is given above the plots. The stacked histogram is split into the following components: \glufu (light pink) and EW (dark pink) \ZZ production, instrumental \ptmiss background (purple), nonresonant proceses (gray), the $\qqZZ$ (blue) and $\qqWZ$ (green) processes, and \tZXtxt production, where $\X$ refers to any other particle.
Postfit refers to individual fits of the data (shown as black points with error bars as uncertainties at 68\%~\CL{}) to the combined $2\ell2\nu+4\ell$ sample, including the \WZ control region, and assuming either SM \Hboson parameters (stacked histogram with the hashed band as the total postfit uncertainty at 68\%~\CL{}) or no \offshell \Hboson production (dot-dashed gold line).
The middle panels along the vertical show the ratio of the data or dashed histograms to the stacked histogram, and the lower panels show the predicted relative contributions of each process.
The rightmost bins contain the overflow.
}
\label{fig:supp:mTZZ-SR}
\end{figure*}

\ifthenelse{\boolean{cms@external}}{
\clearpage
}{}

\begin{figure*}[htbp]
\centering
\includegraphics[width=.3\textwidth]{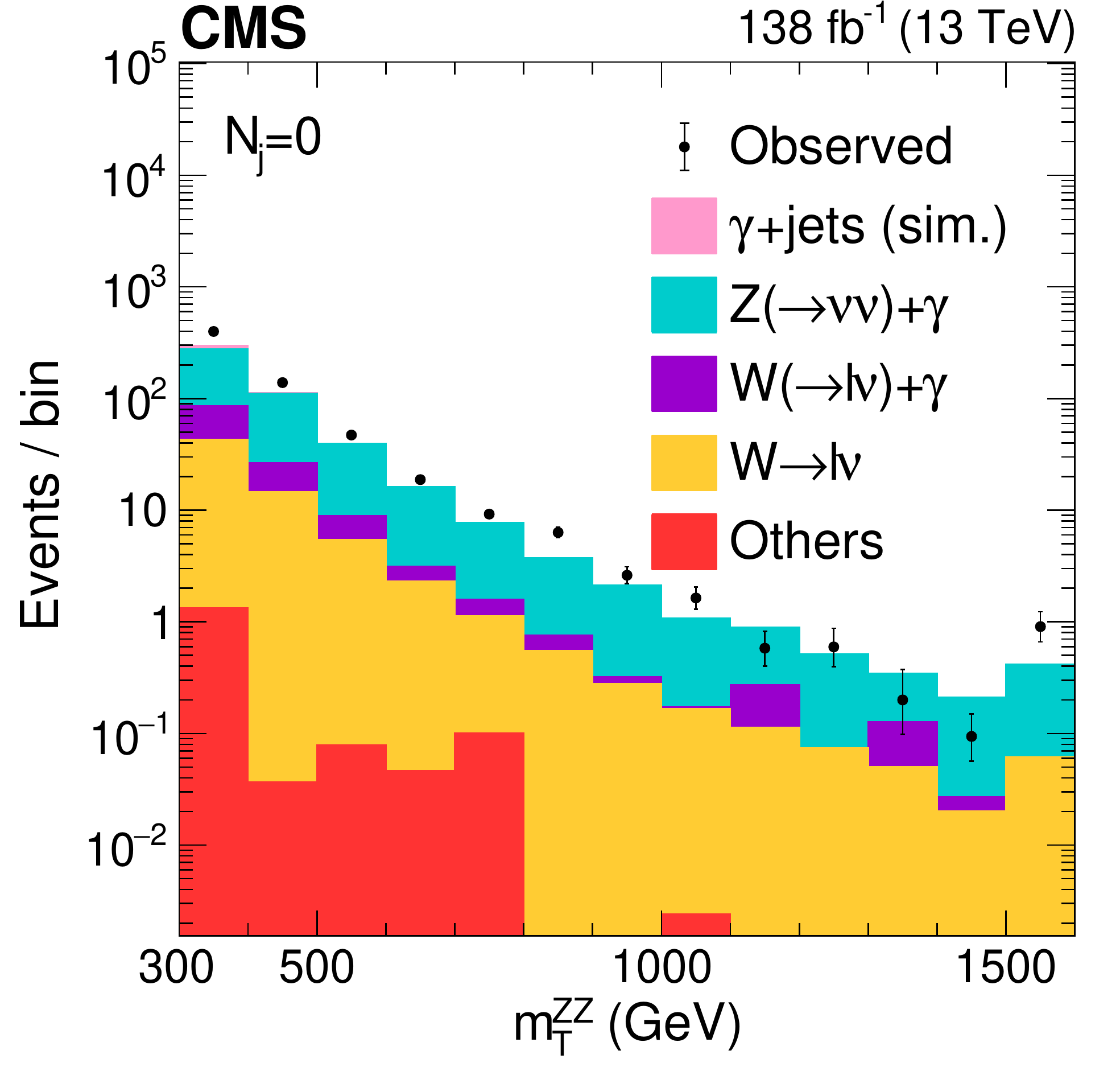}
\includegraphics[width=.3\textwidth]{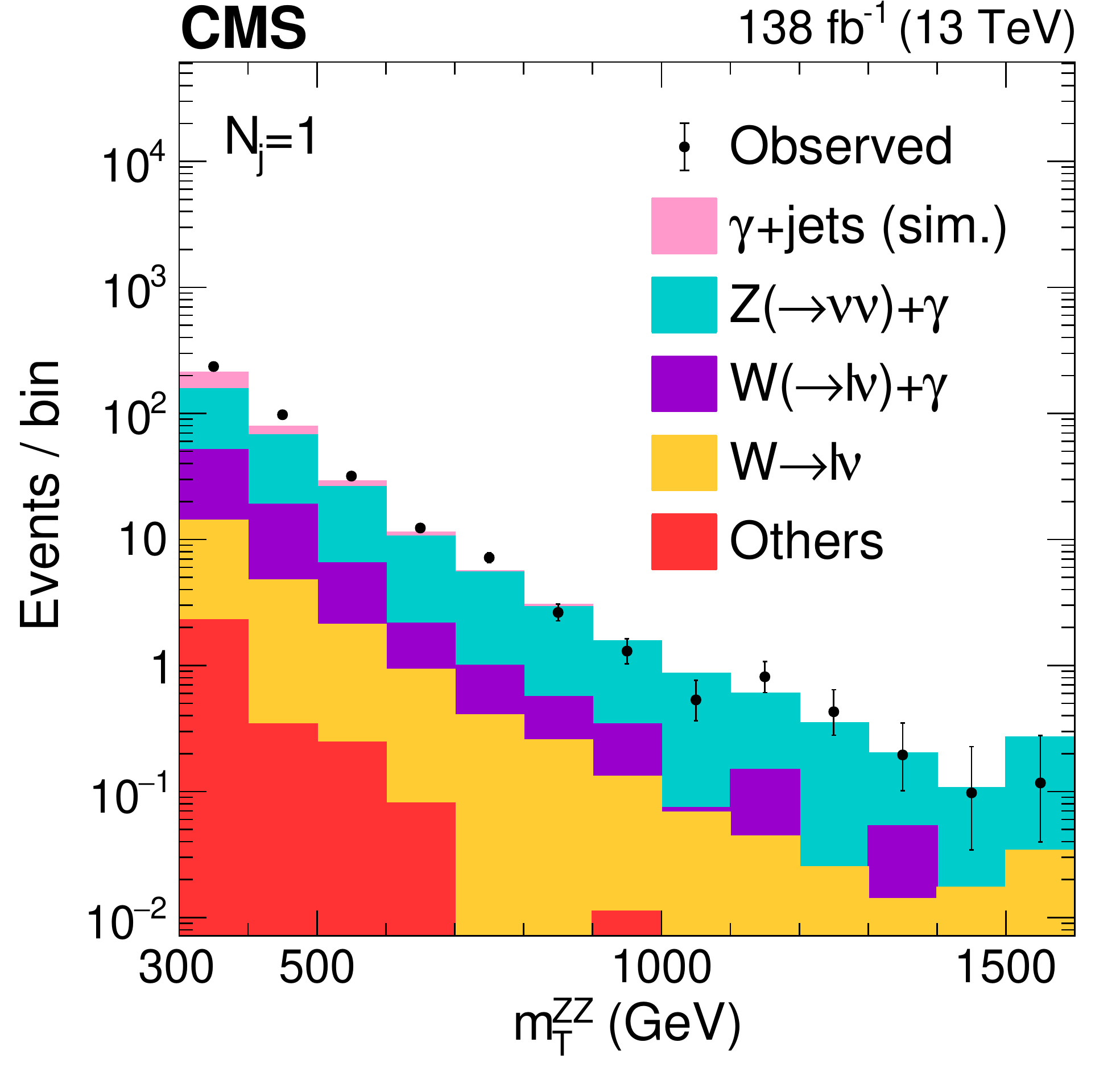}
\includegraphics[width=.3\textwidth]{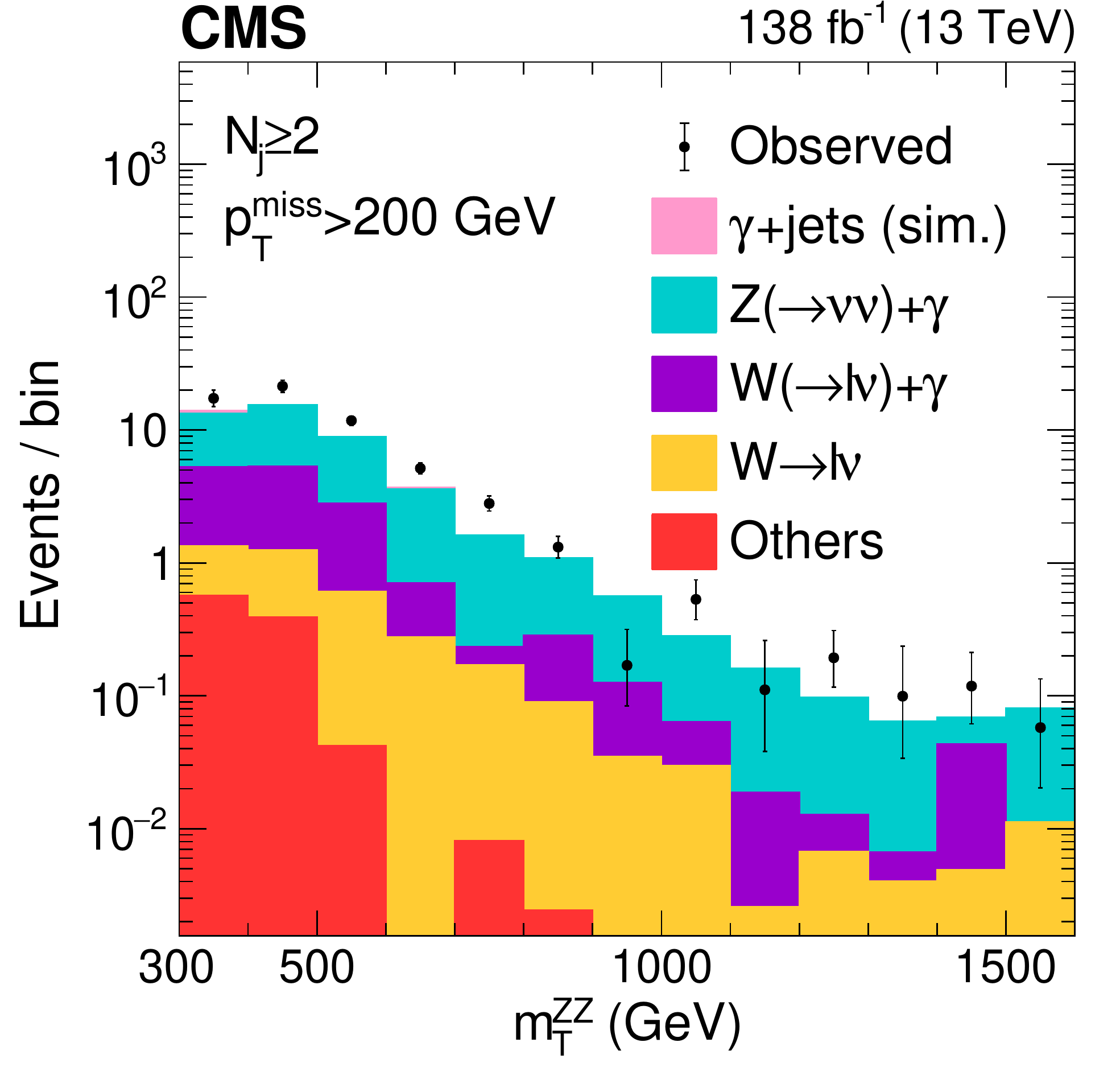}
\caption{
\textbf{Distributions of \mTZZ in the different $\Nj$ categories of the \Gamjetstxt CR.}
The distributions of the transverse \ZZ invariant mass are displayed for the $\Nj=0$, $\Nj=1$, and $\Nj\geq2$ jet multiplicity categories from left to right. The missing transverse momentum requirement $\ptmiss > 200\GeV$ is applied in the $\Nj\geq2$ category to focus on the region more sensitive to \offshell \Hboson production. The stacked histogram shows the predictions for contributions with genuine, large $\ptmiss$, or the instrumental \ptmiss background from the \Gamjetstxt simulation. Contributions with genuine, large $\ptmiss$ are split as those coming from the more dominant \ZnunuGam (teal), \WlnuGam (purple), and \Wlnujetstxt (yellow) processes, and other small components (red). The prediction for instrumental \ptmiss background from simulation is shown in light pink.
The black points with error bars as uncertainties at 68\%~\CL show the observed CR data. The distributions are reweighted with the $\gamma \to \ell\ell$ transfer factors extracted from the $\ptmiss<125\GeV$ sidebands. The rightmost bins include the overflow. In these distributions, we find a discrepancy between the observed data and the predicted distributions
because the reweighted \Gamjetstxt samples have inaccurate $\ptmiss$ response and the simulation is at LO in QCD.
Therefore, we use the difference between the observed data and the genuine-$\ptmiss$ contributions to model the
instrumental $\ptmiss$ background instead of using simulation for this estimate.
}
\label{fig:supp:InstrMET}
\end{figure*}

\ifthenelse{\boolean{cms@external}}{
\clearpage
}{}

\begin{figure*}[htbp]
\centering
\includegraphics[width=.45\textwidth]{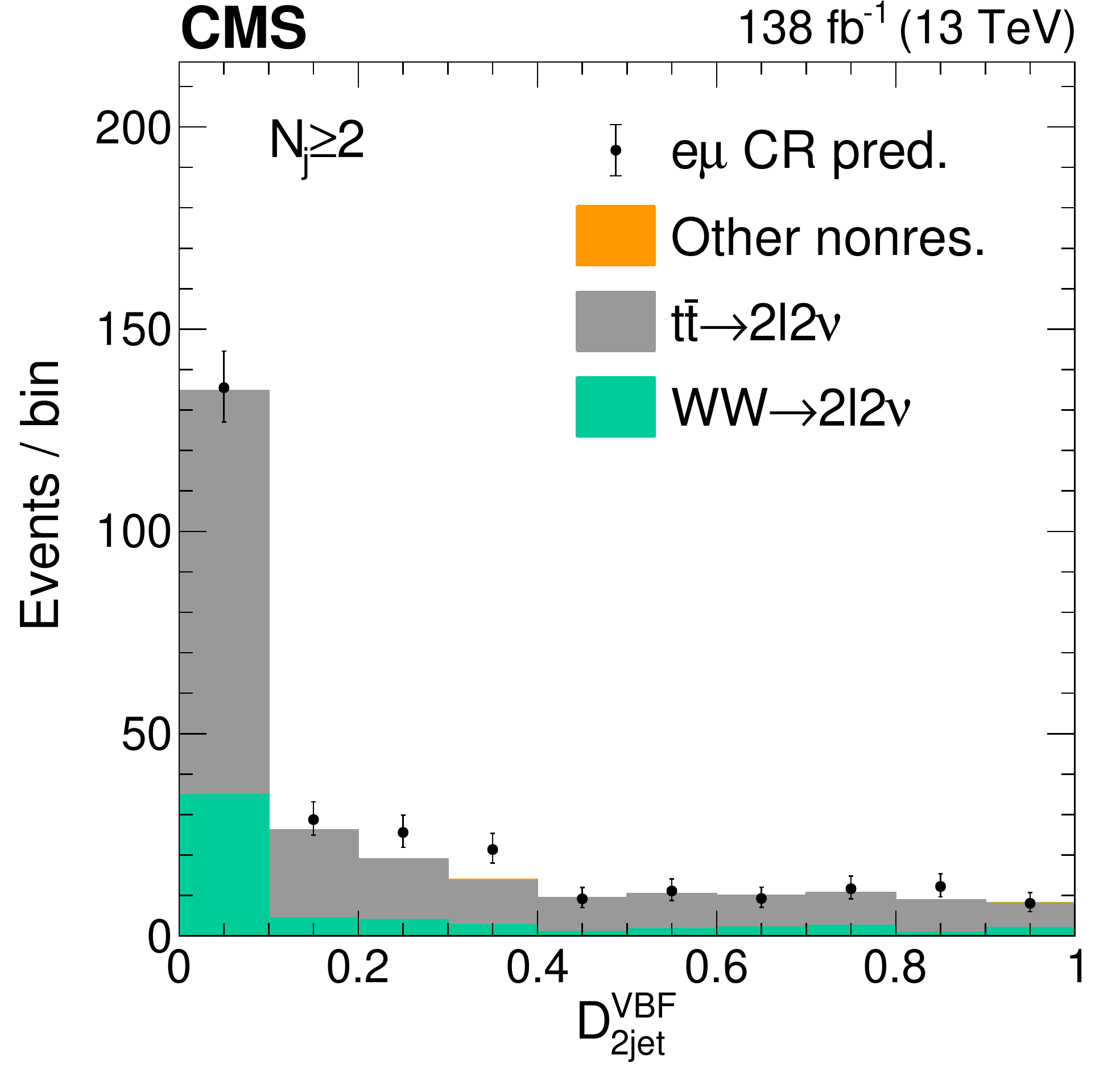}
\includegraphics[width=.45\textwidth]{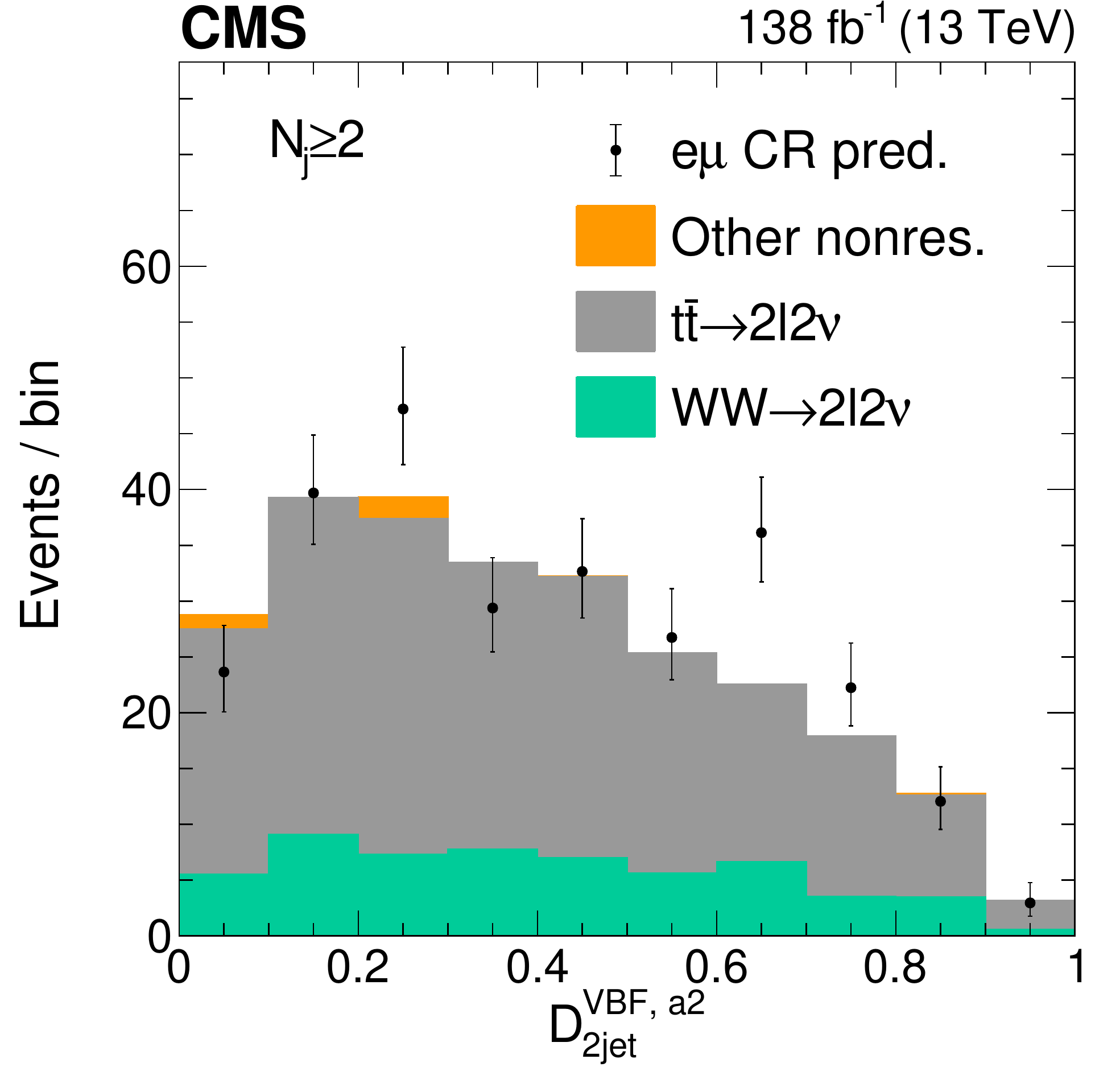}
\caption{
\textbf{Distributions of the \VBF discriminants for nonresonant background.}
The distributions of the SM \DjjVBF (left) and $\DjjVBFHS$ (right) kinematic \VBF discriminants are shown in the $2\ell2\nu$ signal region, $\Nj\geq2$ category. The stacked histogram shows the predictions from simulation, which consists of nonresonant contributions from $\WW$ (green) and $\ttbar$ (gray) production, or other small components (orange). The black points with error bars as uncertainties at 68\%~\CL show the prediction from the $\cPe\cPm$ CR data. While only the data is used in the final estimate of the nonresonant background, we note that predictions from simulation already agree well with the data estimate.
}
\label{fig:supp:NRB}
\end{figure*}

\ifthenelse{\boolean{cms@external}}{
\clearpage
}{}

\begin{figure*}[htbp]
\centering
\includegraphics[width=.3\textwidth]{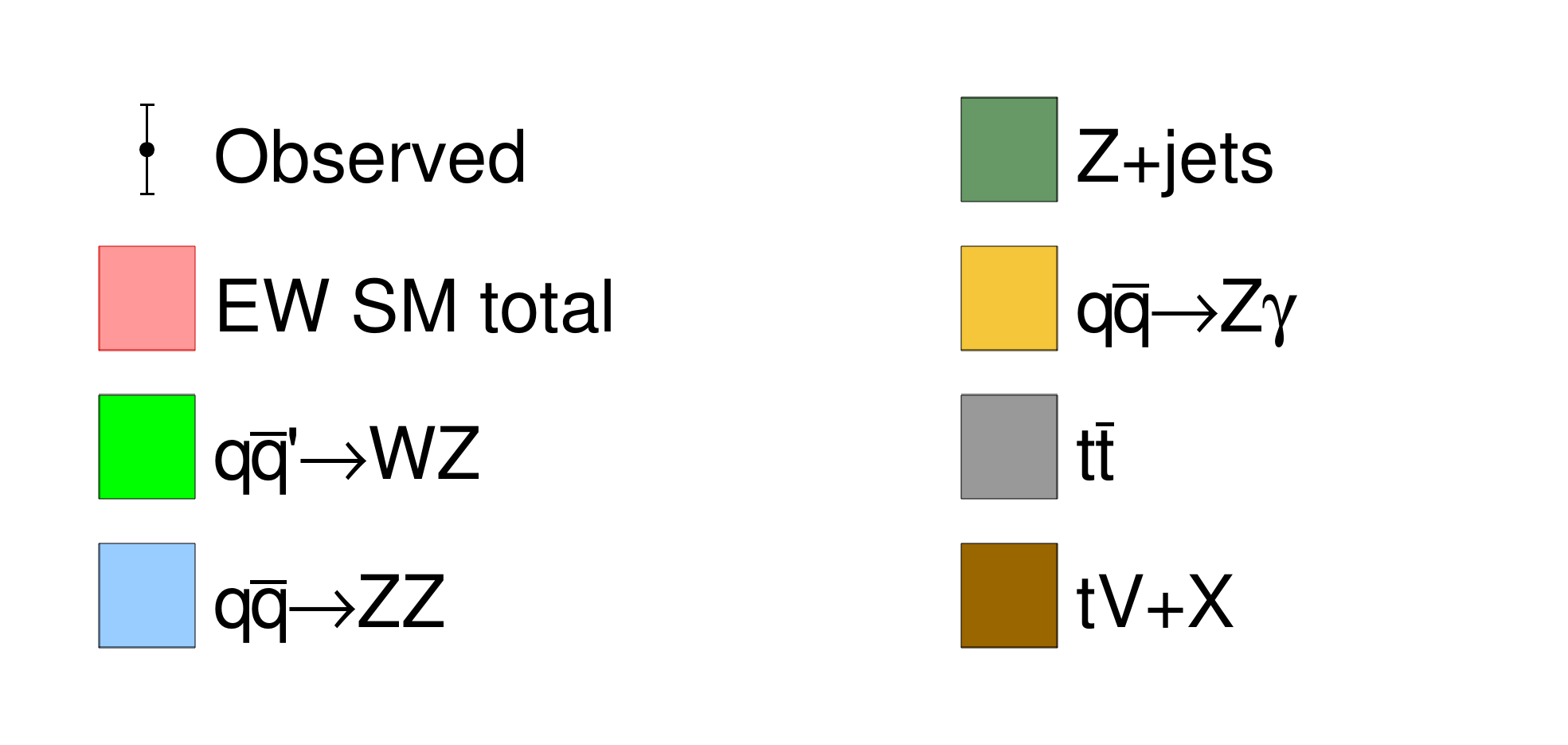} \\
\includegraphics[width=.3\textwidth]{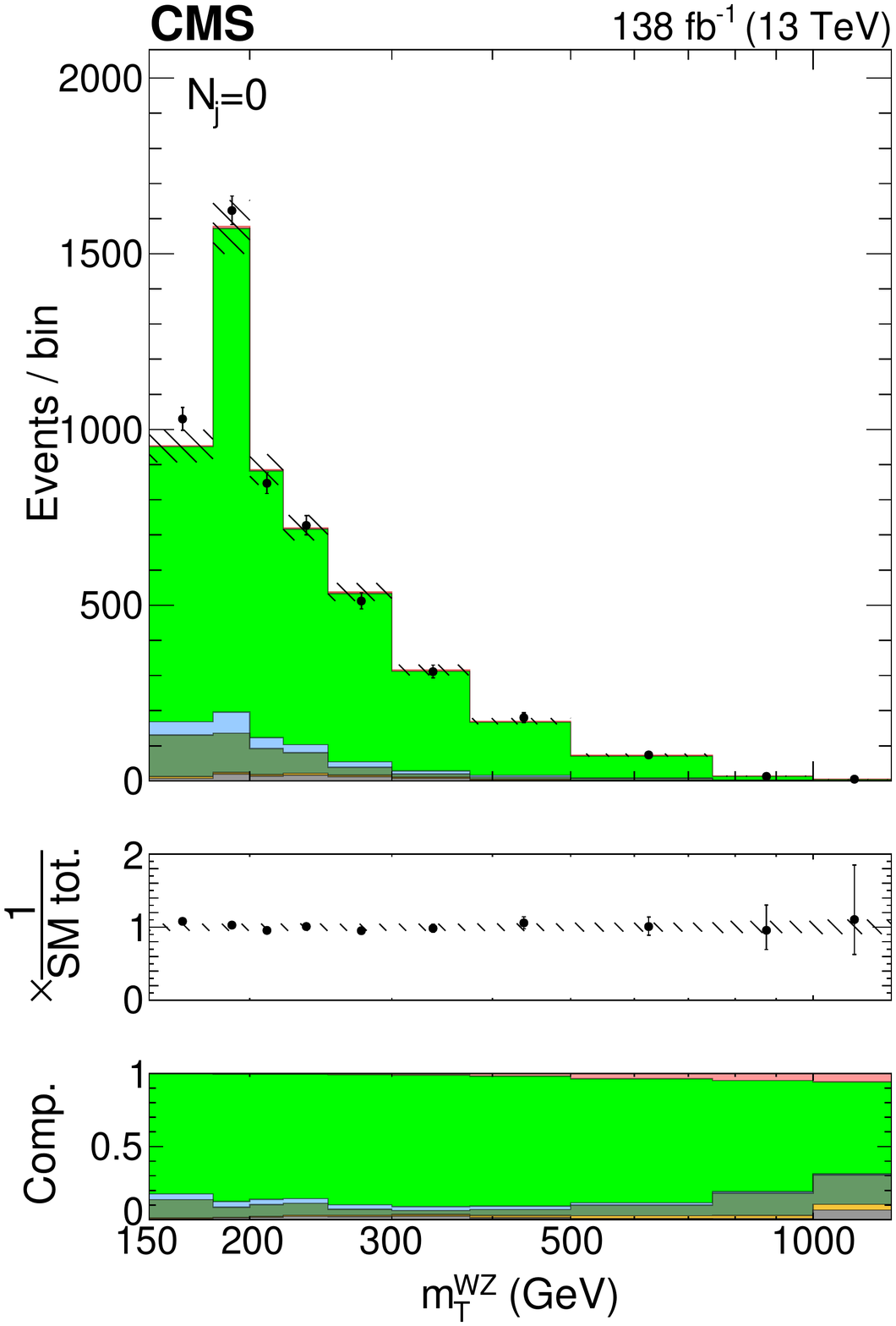}
\includegraphics[width=.3\textwidth]{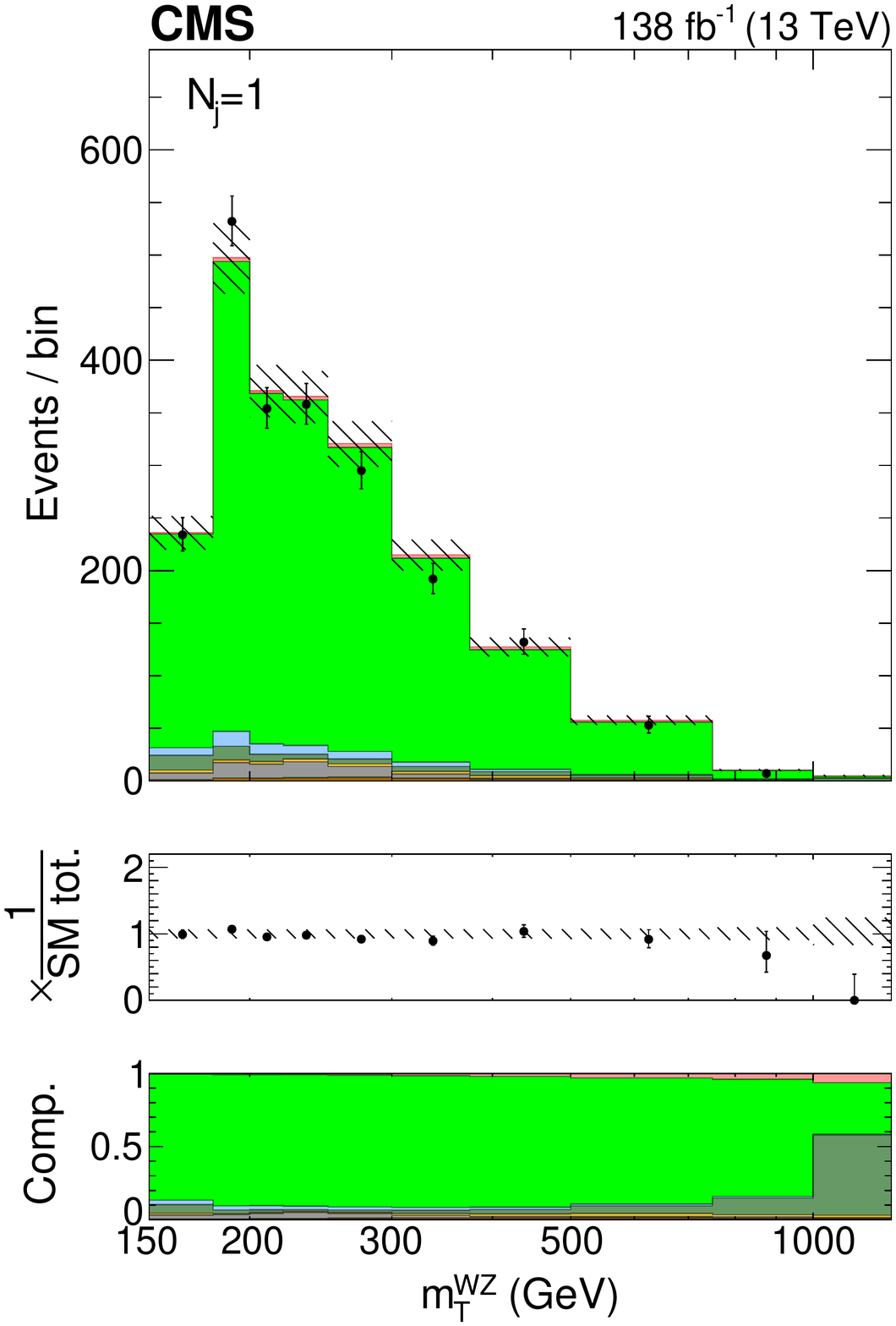}
\includegraphics[width=.3\textwidth]{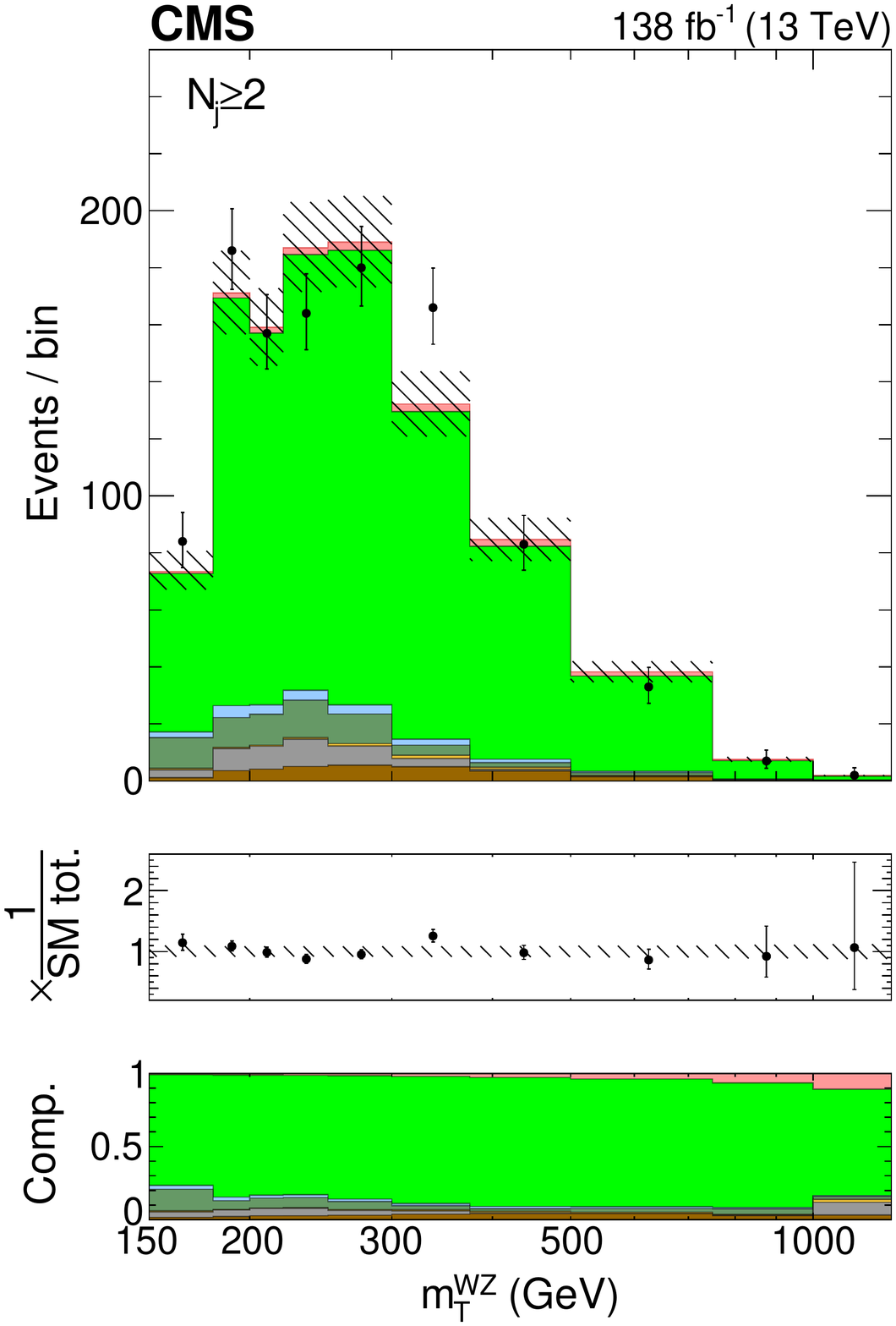}
\caption{
\textbf{Distributions of \mTWZ in different $\Nj$ categories of the $\WZ$ control region.}
The postfit distributions of the transverse \WZ invariant mass are displayed for the $\Nj=0$, $\Nj=1$, and $\Nj\geq2$ jet multiplicity categories of the $\WZ \to 3\ell1\nu$ control region from left to right.
Postfit refers to a combined $2\ell2\nu+4\ell$ fit, together with this control region, assuming SM \Hboson parameters.
The stacked histogram is shown with the hashed band as the total postfit uncertainty at 68\%~\CL{}.
The color legend is given above the plots, with the different contributions referring to the \WZ (light green), \ZZ (blue), \Zjetstxt (dark green), $\ZGam$ (yellow), $\ttbar$ (gray), and \tVXtxt (brown, with $\X$ being any other particle) production processes, as well as the small EW \ZZ production component (dark pink).
The black points with error bars as uncertainties at 68\%~\CL show the observed data.
The middle panels along the vertical show the ratio of the data to the total prediction, and the lower panels show the predicted relative contributions of each process.
The rightmost bins contain the overflow.
}
\label{fig:supp:WZCR}
\end{figure*}

\ifthenelse{\boolean{cms@external}}{
\clearpage
}{}

\begin{figure*}[htbp]
\centering
\includegraphics[width=0.3\textwidth]{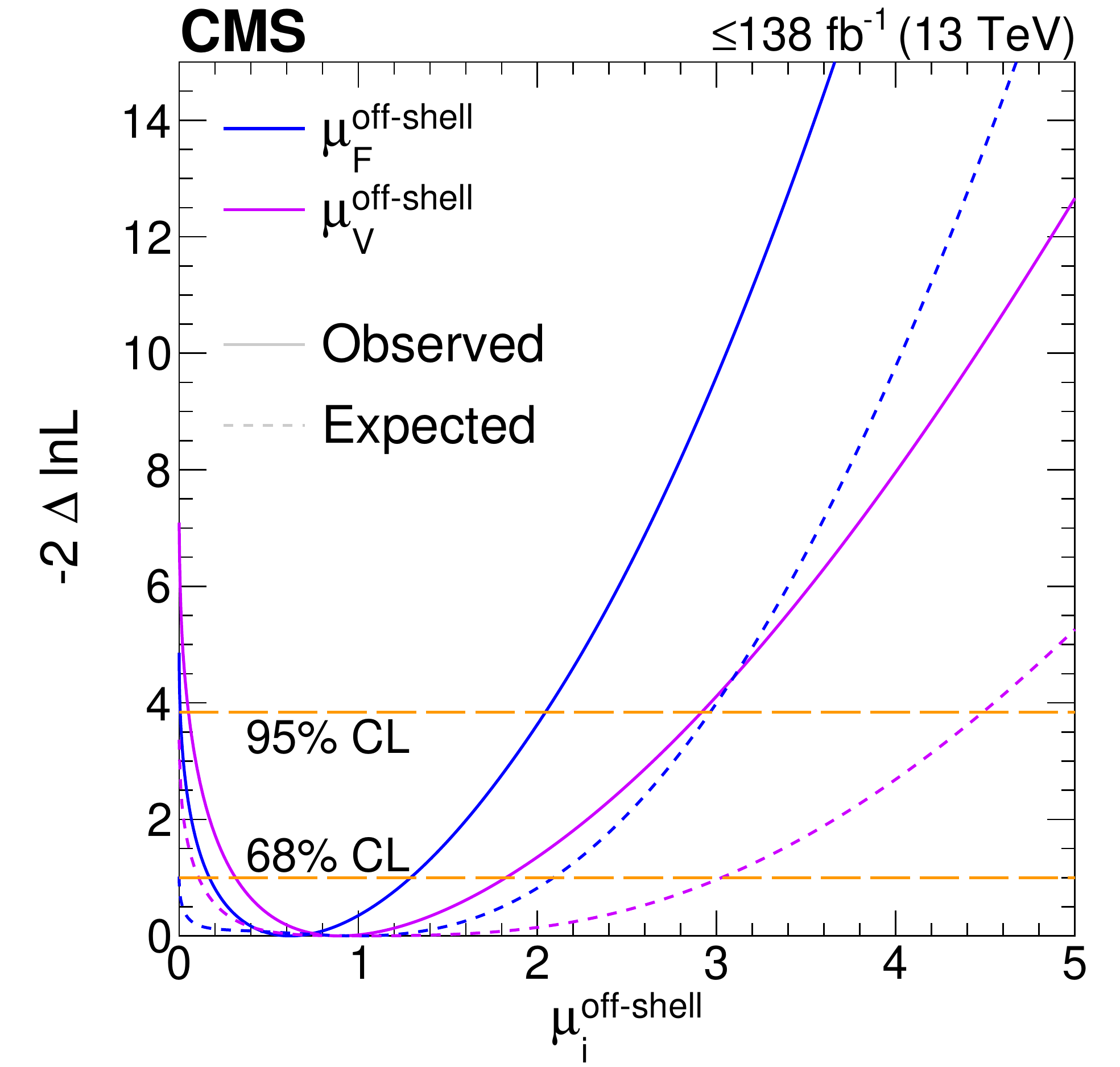}
\includegraphics[width=0.3\textwidth]{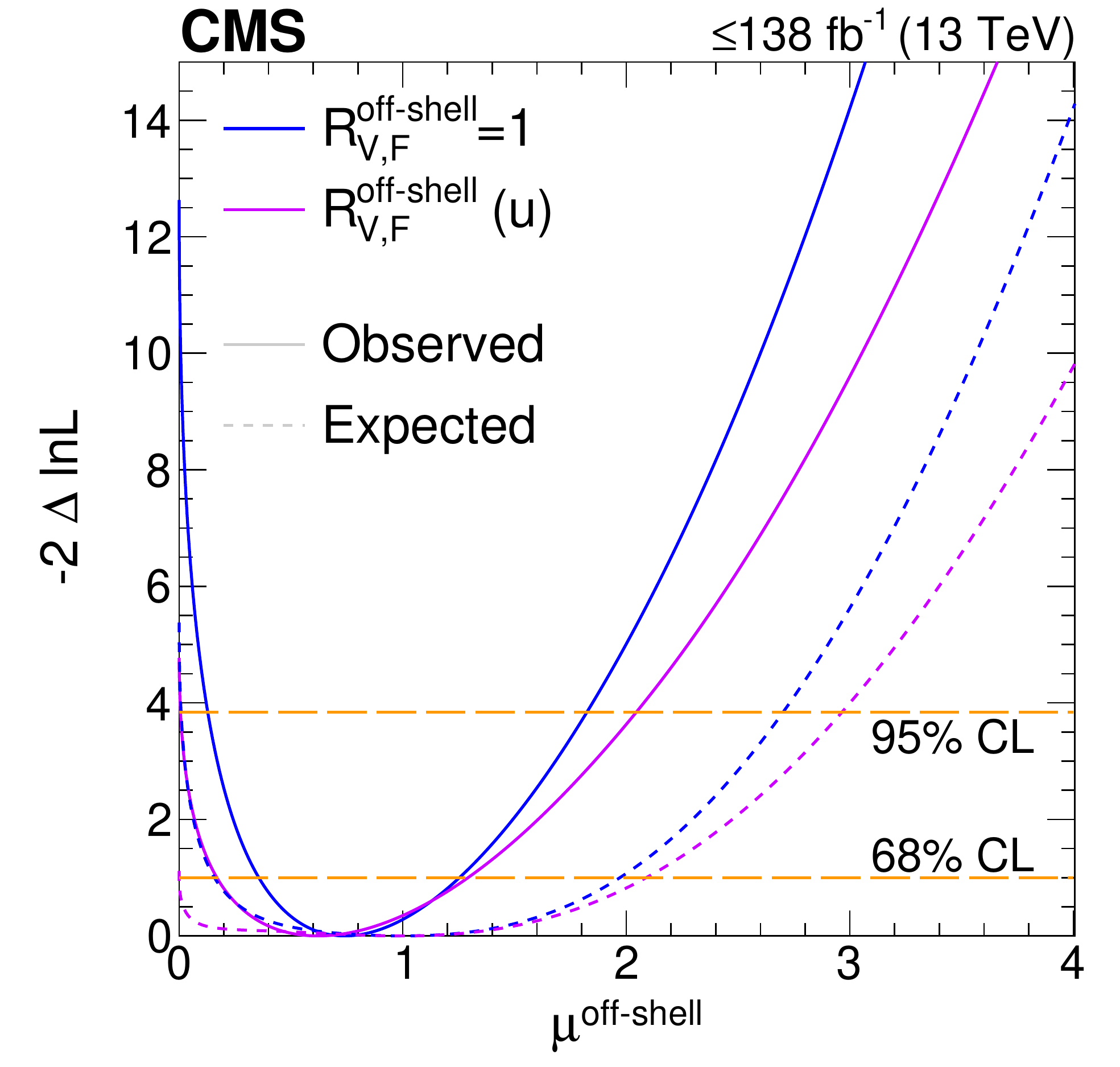}
\includegraphics[width=0.3\textwidth]{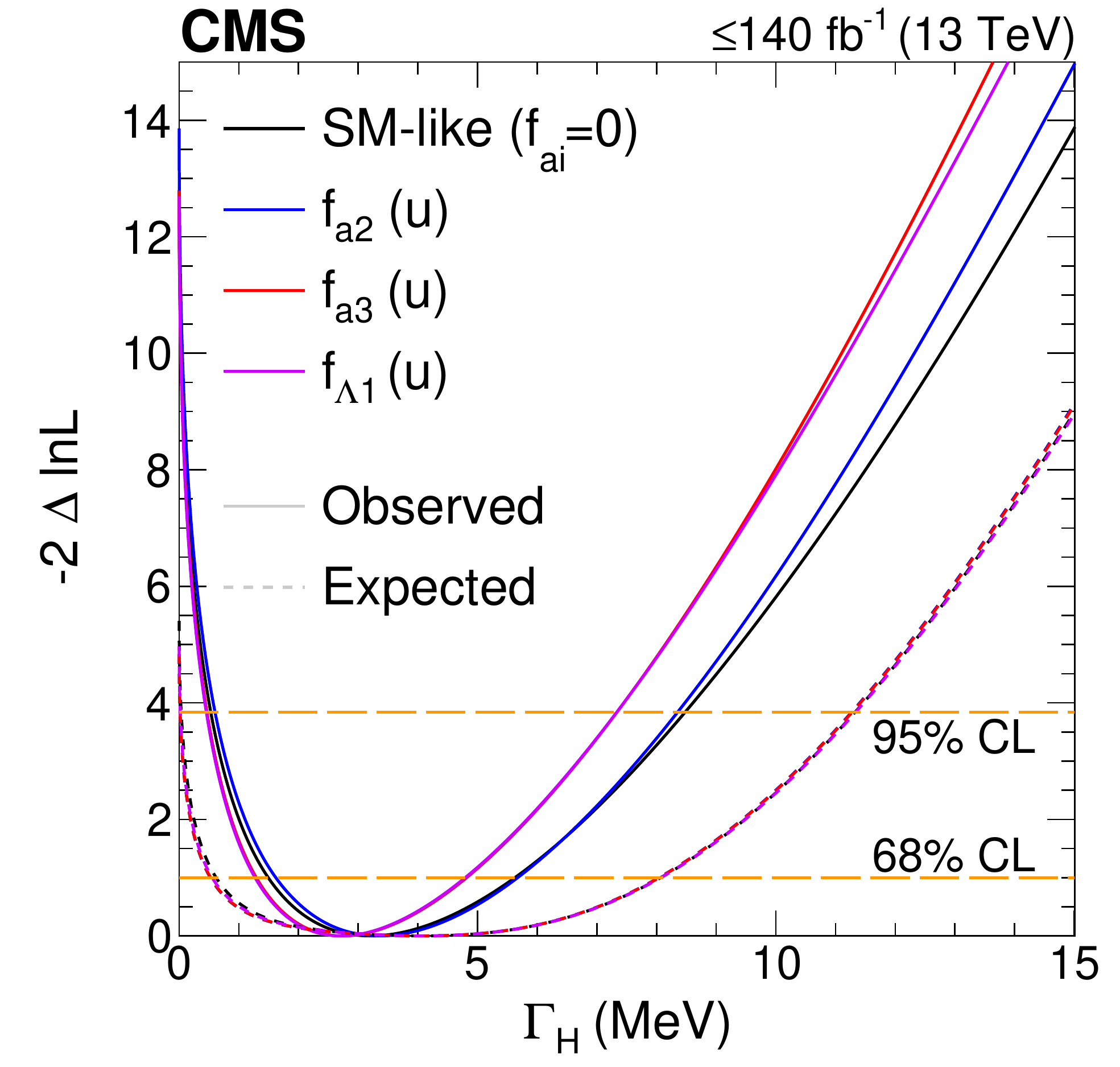}
\\
\includegraphics[width=0.3\textwidth]{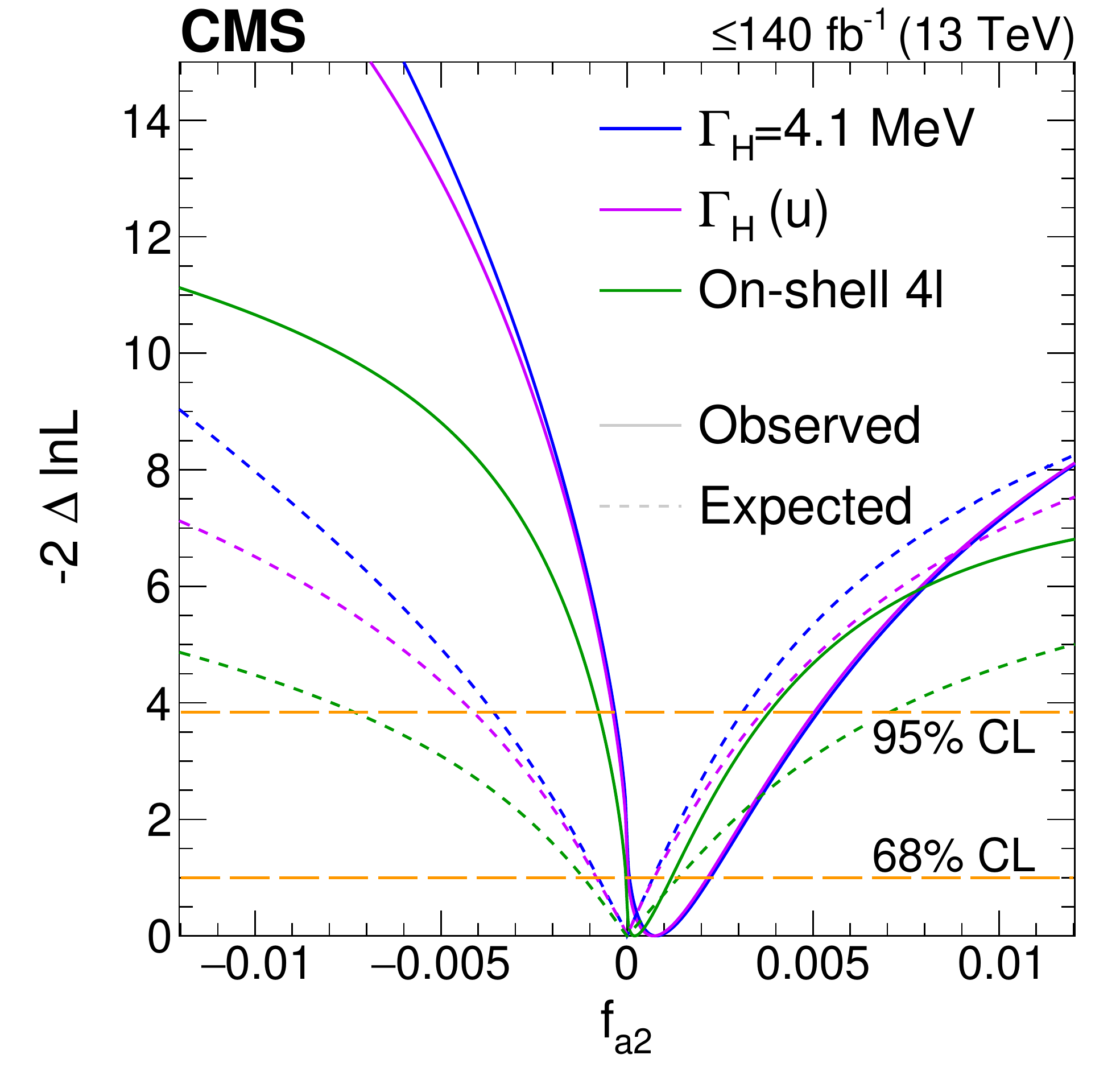}
\includegraphics[width=0.3\textwidth]{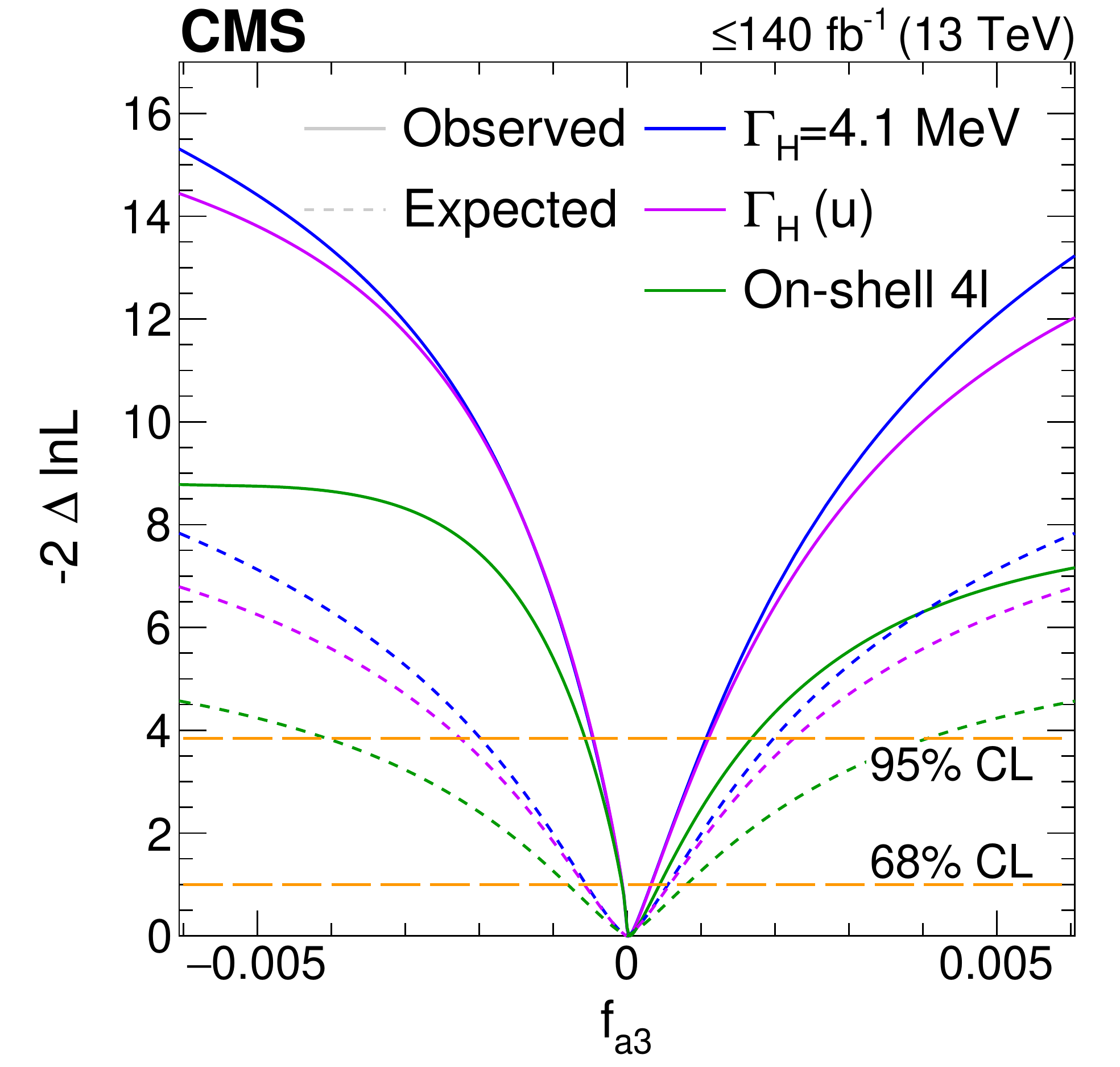}
\includegraphics[width=0.3\textwidth]{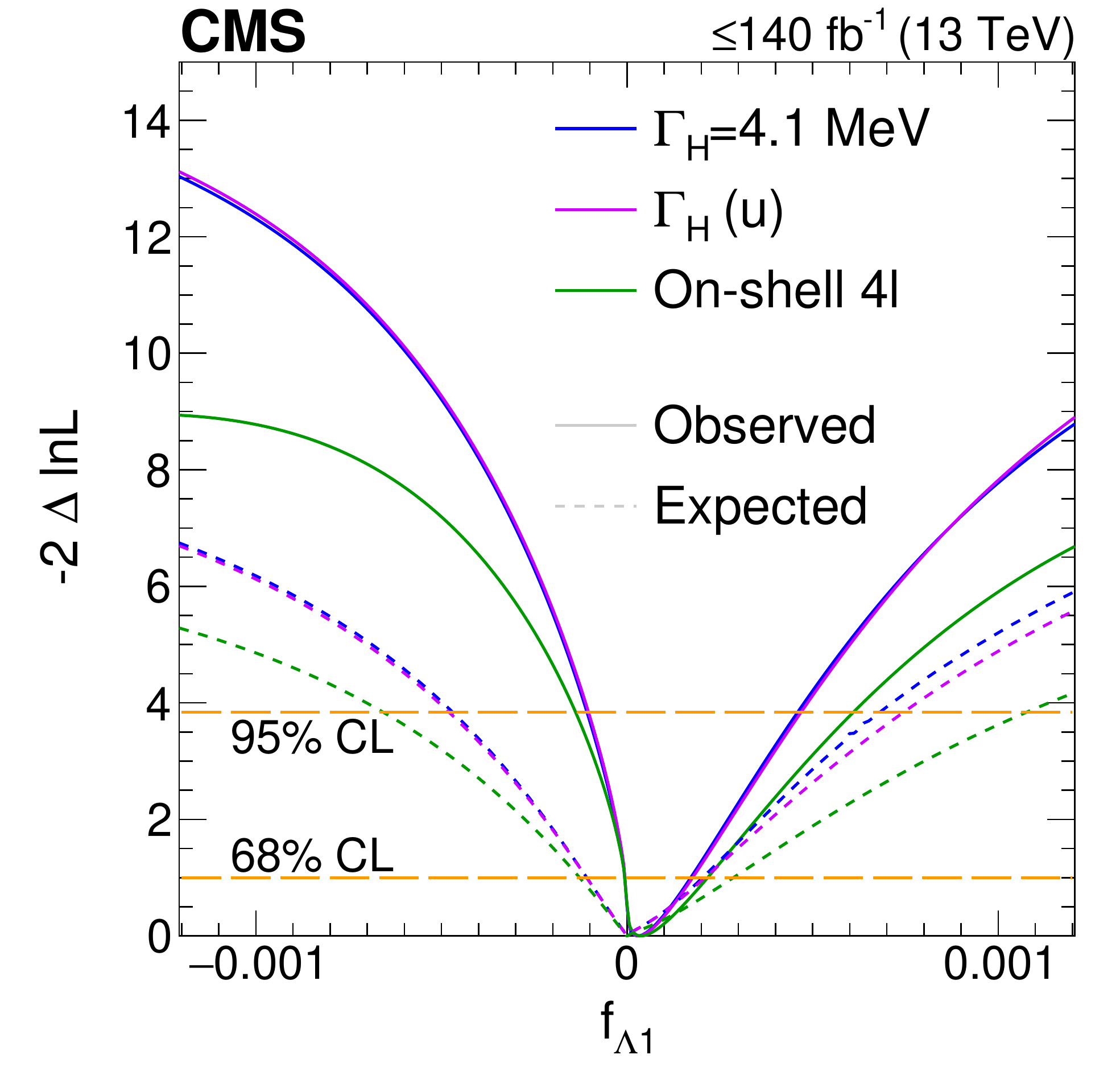}
\caption{
\textbf{Log-likelihood scans of the \offshell signal strengths, $\GH$, and $\fcospai$.}
Top panels: The likelihood scans are shown for $\muFoffsh$ or $\muVoffsh$ (left), $\muoffsh$ (middle), and \GH (right).
Scans for $\muFoffsh$ (blue) and $\muVoffsh$ (magenta) are obtained with the other parameter unconstrained.
Those for $\muoffsh$ are shown with (blue) and without (magenta) the constraint $\RVFoffsh (=\muVoffsh/\muFoffsh) =1$.
Constraints on \GH are shown with and without anomalous \HVV couplings.
Bottom panels: The likelihood scans of the anomalous \HVV coupling parameters \fcospAC{2} (left), \fcospAC{3} (middle), and \fcospLC{1}
(right) are shown with the constraint $\GH=\GHSM=4.1\MeV$ (blue),
\GH unconstrained (magenta), or based on \onshell $4\ell$ data only (green).
Observed (expected) scans are shown with solid (dashed) curves.
The horizontal lines indicate the 68\% ($\dNLL=1.0$) and 95\% ($\dNLL=3.84$)~\CL regions.
The integrated luminosity reaches up to $\lumiDEF$ when only \offshell information is used, and up to $\lumiCDEF$ when \onshell $4\ell$ events are included.
}
\label{fig:supp:interp}
\end{figure*}

\ifthenelse{\boolean{cms@external}}{
\clearpage
}{}

\begin{figure*}[htbp]
\centering
\includegraphics[width=0.45\textwidth]{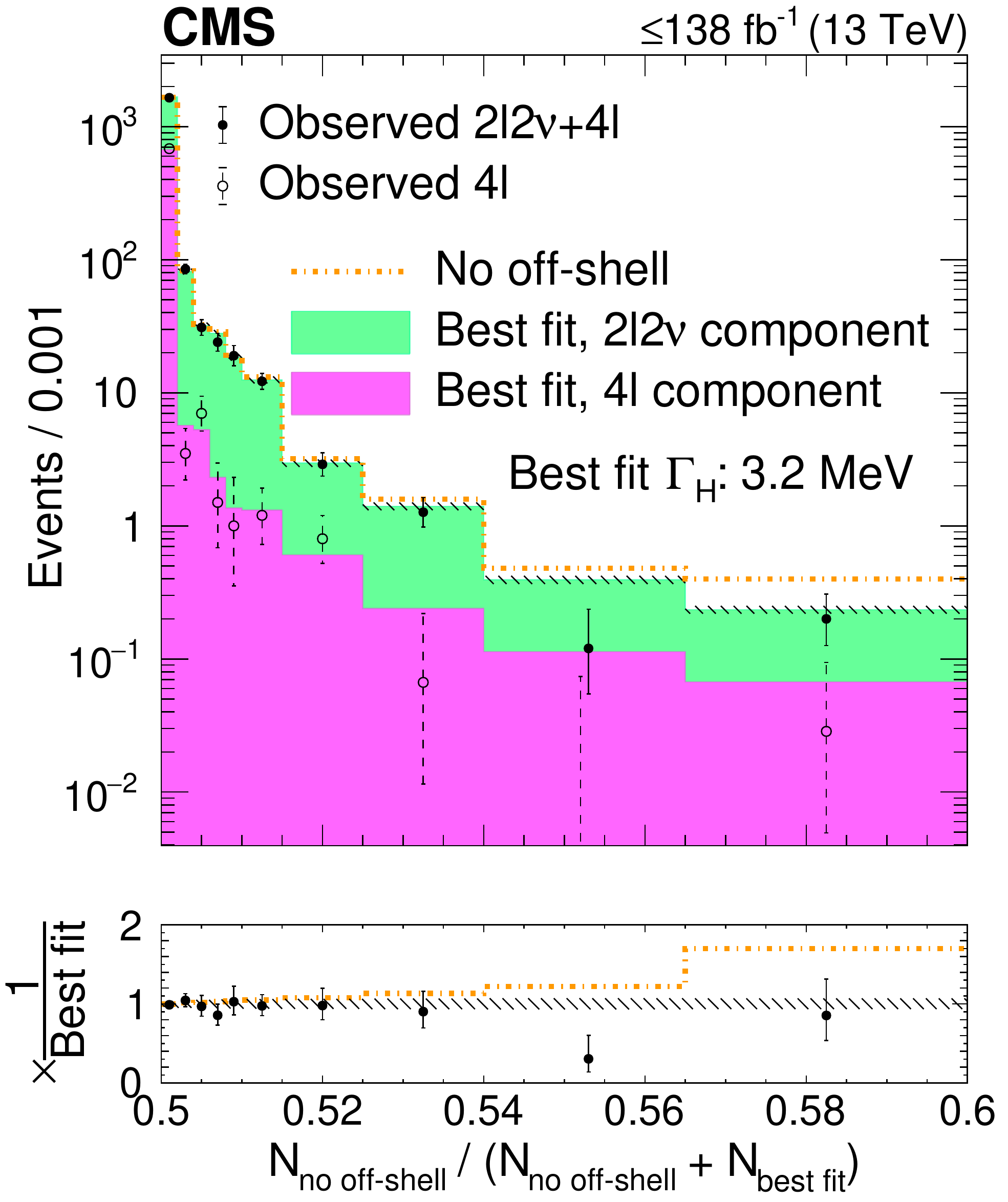}
\caption{
\textbf{Distributions of ratios of the numbers of events in each \offshell signal region bin.}
The ratios are taken after separate fits to the no \offshell hypothesis ($N_{\text{no~\offshell}}$) and the best overall fit ($N_{\text{best~fit}}$) with the observed $\GH$ value of $3.2\MeV$ in the SM-like \HVV couplings scenario.
The stacked histogram displays the predicted contributions (pink from the $4\ell$ \offshell and green from the $2\ell2\nu$ \offshell signal regions) after the best fit, with the hashed band representing the total postfit uncertainty at 68\%~\CL{}, and the gold dot-dashed line shows the predicted distribution of these ratios for a fit to the no \offshell hypothesis. The black solid (hollow) points, with error bars as uncertainties at 68\%~\CL{}, represent the observed $2\ell2\nu$ and $4\ell$ ($4\ell$-only) data. The first and last bins contain the underflow and the overflow, respectively.
The bottom panel displays the ratio of the various displayed hypotheses or observed data to the prediction from the best fit.
The integrated luminosity reaches only up to $\lumiDEF$ since \onshell $4\ell$ events are not displayed.
}
\label{fig:supp:HypoLL}
\end{figure*}

\ifthenelse{\boolean{cms@external}}{
\clearpage
}{}

\begin{table*}[htbp]
\centering
\topcaption
{
\textbf{Results on $\GH$ and the different anomalous \HVV couplings.}
The results on $\GH$ are displayed in units of {\MeV{}}, and those on the anomalous \HVV couplings are summarized in terms of the corresponding \onshell cross section fractions $\fAC{2}$, \fAC{3}, and \fLC{1} ($\fcospai$ in short, and scaled by $10^5$).
For the results on $\GH$, the tests with the anomalous \HVV couplings are distinguished by the denoted $\fcospai$, and the expected best-fit values, not quoted explicitly in the table, are always $\GH=4.1\MeV$.
The SM-like result is the same as that from the combination of all $4\ell$ and $2\ell2\nu$ data sets in Table~\ref{table:interp}.
For the results on $\fcospai$, the constraints are shown with either $\GH=\GHSM=4.1\MeV$ required, or $\GH$ left unconstrained, and the expected best-fit values, also not quoted explicitly, are always null.
The various fit conditions are indicated in the column labeled ``Condition'', where the abbreviation ``(u)'' indicates which parameter is unconstrained.
}
\renewcommand{\arraystretch}{1.25}
\begin{tabular}{ccccccc}
\multirow{2}{*}{Parameter} & \multirow{2}{*}{Condition}
 & \multicolumn{3}{c}{Observed} & \multicolumn{2}{c}{Expected}  \\
 & & Best fit & $68\%$~\CL & $95\%$~\CL & $68\%$~\CL & $95\%$~\CL  \\
\hline
\multirow{4}{*}{\GH ($\MeVns$)}
 & SM-like            
 & $3.2$ & $[1.5, 5.6]$ & $[0.5, 8.5]$ & $[0.6, 8.1]$ & $[0.03, 11.3]$ \\
 & \fcospAC{2} (u)
 & $3.4$ & $[1.6, 5.7]$ & $[0.6, 8.4]$ & $[0.5, 8.0]$ & $[0.02, 11.3]$ \\
 & \fcospAC{3} (u)
 & $2.7$ & $[1.3, 4.8]$ & $[0.5, 7.3]$ & $[0.5, 8.0]$ & $[0.02, 11.3]$ \\
 & \fcospLC{1} (u)
 & $2.7$ & $[1.3, 4.8]$ & $[0.5, 7.3]$ & $[0.6, 8.1]$ & $[0.02, 11.3]$ \\
\noalign{\vskip 3mm}
\multirow{2}{*}{$\fcospAC{2}\, (\times 10^5)$}
  & $\GH=\GHSM$
  & $79$ & $[6.6, 225]$ & $[-32, 514]$ & $[-78, 70]$ & $[-359, 311]$ \\
  & \GH (u)
  & $72$ & $[2.7, 216]$ & $[-38, 503]$ & $[-82, 73]$ & $[-413, 364]$ \\
\noalign{\vskip 3mm}
\multirow{2}{*}{$\fcospAC{3}\, (\times 10^5)$}
  & $\GH=\GHSM$
  & $2.2$ & $[-6.4, 32]$ & $[-46, 107]$ & $[-55, 55]$ & $[-198, 198]$ \\
  & \GH (u)
  & $2.4$ & $[-6.2, 33]$ & $[-46, 110]$ & $[-58, 58]$ & $[-225, 225]$ \\
\noalign{\vskip 3mm}
\multirow{2}{*}{$\fcospLC{1}\, (\times 10^5)$}
  & $\GH=\GHSM$
  & $2.9$ & $[-0.62, 17]$ & $[-11, 46]$ & $[-11, 20]$ & $[-47, 68]$ \\
  & \GH (u)
  & $3.1$ & $[-0.56, 18]$ & $[-10, 47]$ & $[-11, 21]$ & $[-48, 75]$ \\
\end{tabular}
\label{table:supp:interp}
\end{table*}

}

\clearpage

{

\ifthenelse{\boolean{cms@external}}{
}{
\renewcommand{\sectionmark}[1]{\markright{\textbf{#1}}}
\sectionmark{Acknowledgments}
}

\begin{acknowledgments}
  We congratulate our colleagues in the CERN accelerator departments for the excellent performance of the LHC and thank the technical and administrative staffs at CERN and at other CMS institutes for their contributions to the success of the CMS effort. In addition, we gratefully acknowledge the computing centers and personnel of the Worldwide LHC Computing Grid and other centers for delivering so effectively the computing infrastructure essential to our analyses. Finally, we acknowledge the enduring support for the construction and operation of the LHC, the CMS detector, and the supporting computing infrastructure provided by the following funding agencies: BMBWF and FWF (Austria); FNRS and FWO (Belgium); CNPq, CAPES, FAPERJ, FAPERGS, and FAPESP (Brazil); MES and BNSF (Bulgaria); CERN; CAS, MoST, and NSFC (China); MINCIENCIAS (Colombia); MSES and CSF (Croatia); RIF (Cyprus); SENESCYT (Ecuador); MoER, ERC PUT and ERDF (Estonia); Academy of Finland, MEC, and HIP (Finland); CEA and CNRS/IN2P3 (France); BMBF, DFG, and HGF (Germany); GSRI (Greece); NKFIA (Hungary); DAE and DST (India); IPM (Iran); SFI (Ireland); INFN (Italy); MSIP and NRF (Republic of Korea); MES (Latvia); LAS (Lithuania); MOE and UM (Malaysia); BUAP, CINVESTAV, CONACYT, LNS, SEP, and UASLP-FAI (Mexico); MOS (Montenegro); MBIE (New Zealand); PAEC (Pakistan); MSHE and NSC (Poland); FCT (Portugal); JINR (Dubna); MON, RosAtom, RAS, RFBR, and NRC KI (Russia); MESTD (Serbia); MCIN/AEI and PCTI (Spain); MOSTR (Sri Lanka); Swiss Funding Agencies (Switzerland); MST (Taipei); ThEPCenter, IPST, STAR, and NSTDA (Thailand); TUBITAK and TAEK (Turkey); NASU (Ukraine); STFC (United Kingdom); DOE and NSF (USA).

\hyphenation{Rachada-pisek} Individuals have received support from the Marie-Curie program and the European Research Council and Horizon 2020 Grant, contract Nos.\ 675440, 724704, 752730, 758316, 765710, 824093, 884104, and COST Action CA16108 (European Union); the Leventis Foundation; the Alfred P.\ Sloan Foundation; the Alexander von Humboldt Foundation; the Belgian Federal Science Policy Office; the Fonds pour la Formation \`a la Recherche dans l'Industrie et dans l'Agriculture (FRIA-Belgium); the Agentschap voor Innovatie door Wetenschap en Technologie (IWT-Belgium); the F.R.S.-FNRS and FWO (Belgium) under the ``Excellence of Science -- EOS" -- be.h project n.\ 30820817; the Beijing Municipal Science \& Technology Commission, No. Z191100007219010; the Ministry of Education, Youth and Sports (MEYS) of the Czech Republic; the Deutsche Forschungsgemeinschaft (DFG), under Germany's Excellence Strategy -- EXC 2121 ``Quantum Universe" -- 390833306, and under project number 400140256 - GRK2497; the Lend\"ulet (``Momentum") Program and the J\'anos Bolyai Research Scholarship of the Hungarian Academy of Sciences, the New National Excellence Program \'UNKP, the NKFIA research grants 123842, 123959, 124845, 124850, 125105, 128713, 128786, and 129058 (Hungary); the Council of Science and Industrial Research, India; the Latvian Council of Science; the Ministry of Science and Higher Education and the National Science Center, contracts Opus 2014/15/B/ST2/03998 and 2015/19/B/ST2/02861 (Poland); the Funda\c{c}\~ao para a Ci\^encia e a Tecnologia, grant CEECIND/01334/2018 (Portugal); the National Priorities Research Program by Qatar National Research Fund; the Ministry of Science and Higher Education, projects no. 0723-2020-0041 and no. FSWW-2020-0008 (Russia); MCIN/AEI/10.13039/501100011033, ERDF ``a way of making Europe", and the Programa Estatal de Fomento de la Investigaci{\'o}n Cient{\'i}fica y T{\'e}cnica de Excelencia Mar\'{\i}a de Maeztu, grant MDM-2017-0765 and Programa Severo Ochoa del Principado de Asturias (Spain); the Stavros Niarchos Foundation (Greece); the Rachadapisek Sompot Fund for Postdoctoral Fellowship, Chulalongkorn University and the Chulalongkorn Academic into Its 2nd Century Project Advancement Project (Thailand); the Kavli Foundation; the Nvidia Corporation; the SuperMicro Corporation; the Welch Foundation, contract C-1845; and the Weston Havens Foundation (USA).
\end{acknowledgments}

}
\cleardoublepage \section{The CMS Collaboration \label{app:collab}}\begin{sloppypar}\hyphenpenalty=5000\widowpenalty=500\clubpenalty=5000\cmsinstitute{Yerevan~Physics~Institute, Yerevan, Armenia}
A.~Tumasyan
\cmsinstitute{Institut~f\"{u}r~Hochenergiephysik, Vienna, Austria}
W.~Adam\cmsorcid{0000-0001-9099-4341}, J.W.~Andrejkovic, T.~Bergauer\cmsorcid{0000-0002-5786-0293}, S.~Chatterjee\cmsorcid{0000-0003-2660-0349}, K.~Damanakis, M.~Dragicevic\cmsorcid{0000-0003-1967-6783}, A.~Escalante~Del~Valle\cmsorcid{0000-0002-9702-6359}, R.~Fr\"{u}hwirth\cmsAuthorMark{1}, M.~Jeitler\cmsAuthorMark{1}\cmsorcid{0000-0002-5141-9560}, N.~Krammer, L.~Lechner\cmsorcid{0000-0002-3065-1141}, D.~Liko, I.~Mikulec, P.~Paulitsch, F.M.~Pitters, J.~Schieck\cmsAuthorMark{1}\cmsorcid{0000-0002-1058-8093}, R.~Sch\"{o}fbeck\cmsorcid{0000-0002-2332-8784}, D.~Schwarz, S.~Templ\cmsorcid{0000-0003-3137-5692}, W.~Waltenberger\cmsorcid{0000-0002-6215-7228}, C.-E.~Wulz\cmsAuthorMark{1}\cmsorcid{0000-0001-9226-5812}
% ...
\vskip\cmsinstskip
% ...
1:~Now at TU Wien, Wien, Austria\\
% ...
\end{sloppypar}
\end{document}